\documentclass[aps,twocolumn,pra,superscriptaddress,eps2pdf]{revtex4-1}

\usepackage{epsfig}
\usepackage{amsfonts}
\usepackage{amssymb}
\usepackage{mathrsfs}
\usepackage{theorem}
\usepackage{amsmath}
\usepackage{times}
\usepackage{color}
\usepackage{soul}

\definecolor{jens}{rgb}{.2,0.7,.9}
\definecolor{earl}{rgb}{.7,0.1,.9}
\usepackage{ifpdf}
\ifpdf
\usepackage{epstopdf}   
\fi

\newtheorem{theorem}{Theorem}
\newtheorem{defin}{Definition}
\newtheorem{lem}{Lemma}
\newtheorem{corollary}{Corollary}

\newcommand{\ket}[1]{\ensuremath{\vert#1\rangle}}
\newcommand{\bra}[1]{\ensuremath{\langle #1\vert}}

\newcommand{\kb}[2]{\ensuremath{\vert #1 \rangle \langle #2 \vert}}

\newcommand{\ox}[0]{\ensuremath{\otimes}}

 \newcommand{\earl}[1]{#1}  
 
\def\R{\mathbb{R}}
\def\C{\mathbb{C}}

\renewcommand{\vec}[1]{\ensuremath{\mathbf{#1}}}

\newcommand{\tr}{\ensuremath{\mathrm{tr}}}

\def\id{\mbox{\small 1} \!\! \mbox{1}}

\def\id{\mbox{\small 1} \!\! \mbox{1}}

\def\id{{\mathchoice {\rm 1\mskip-4mu l} {\rm 1\mskip-4mu l} {\rm 1\mskip-4.5mu l} {\rm 1\mskip-5mu l}}}

\begin{document}
\title{Continuous-variable entanglement distillation and non-commutative central limit theorems}

\author{Earl T.\ Campbell}
\affiliation{Dahlem Center for Complex Quantum Systems, Freie Universit{\"a}t Berlin, 14195 Berlin, Germany}
\email{earltcampbell@gmail.com}
\author{Marco G.\ Genoni}
\affiliation{QOLS, Blackett Laboratory, Imperial College London, London SW7 2BW, UK}
\author{Jens Eisert}
\affiliation{Dahlem Center for Complex Quantum Systems, Freie Universit{\"a}t Berlin, 14195 Berlin, Germany}


\begin{abstract}
Entanglement distillation transforms weakly entangled noisy states into highly entangled states, a primitive
to be used in quantum repeater schemes and other protocols designed for quantum communication and key distribution.
In this work, we present a comprehensive framework for continuous-variable entanglement distillation schemes that 
convert noisy non-Gaussian states into Gaussian ones in many iterations of the protocol. Instances of these protocols
include (a) the recursive-Gaussifier  protocol, (b) the temporally-reordered recursive-Gaussifier protocol, and (c) the 
pumping-Gaussifier protocol. The flexibility of these protocols give rise to
several beneficial trade-offs related to success probabilities or memory requirements, which that can be adjusted to reflect
experimental demands. Despite these protocols involving measurements,
we relate the convergence in this protocols to new instances of non-commutative central limit theorems, in a
formalism that we lay out in great detail. Implications of the findings for quantum repeater schemes are discussed.

\end{abstract}

\maketitle  

\section{Introduction}

Photons, with information encoded in continuous variable degrees of freedom, can travel great distance without significant decoherence.  We can, using beam-splitters, phase-shifters, and detectors,  coherently manipulate photons and make measurements.  Specifically in the continuous-variable regime, brighter sources of light are available than for single photon, discrete, light sources. These features have motivated research into the usefulness of photonic systems for quantum cryptography, communication and distributed quantum information processing~\cite{Kok07,Eisert03}.  Discrete protocols, for finite-dimensional systems with arbitrary quantum control, do not typically have exact analogs but rather cousins in the linear optical setting.  Any two qubit entangled state can be distilled by local operations~\cite{HHH01a}, whereas distillation of entangled Gaussian state using linear optics is impossible~\cite{NoGoFu,Eisert02,Giedke02}.   Soon after these impossibility proofs were obtained, it was discovered that an initially non-Gaussian state could, using only linear optics, enable entanglement distillation~\cite{Browne03,Eisert04,Eisert07}.  The original distillation protocol, which is conditioned on detectors finding no photons, outputs a state that evolves toward a Gaussian.  Over the years, this protocol has inspired several variants that have been found to exhibit the same Gaussification phenomena~\cite{Fiur07,Fiur10}.   Similar ``no-go" results \cite{NoSq} prohibit the distillation of highly squeezed states using only passive linear optics, although with relaxed constraints some proposals are possible~\cite{Heersink06}. 

\earl{Leaving the realm of purely Gaussian operations is essential for entanglement distillation, but unfortunately non-Gaussian operations are much more experimentally challenging.  Therefore, it is desirable to keep non-Gaussian operations to a minimum.  In the aforementioned protocols, and those considered herein, only the initial noisy resource needs to be non-Gaussian.  A source of Gaussian entangled states, such as those emitted by a pumped parametric downconverter, can be probabilistically de-Gaussified by adding or subtracting single photons through the use of single photon detectors and/or sources~\cite{Subtraction1,Subtraction2,Subtraction3,Fiur10,Eisert04,Kim08}.  An additional benefit of de-Gaussification is that it too can increase the entanglement and other figures of merit, such as the teleportation fidelity~\cite{Opatrn,Cochrane,Olivares03,Dell2}.  Some matter systems (e.g. Ref.~\cite{BiPhotons}) also provide a more direct source of non-Gaussian entangled photons.  These are the most experimentally feasible means of non-Gaussian state preparation, but the potential advantage of exploiting more exotic forms of non-Gaussianity has also been considered~\cite{Dell,Fiur10,Dell2,Genoni10}.  The need for non-Gaussian operations extends beyond distillation problems, and they are required to violate locality~\cite{Bell,Banaszek,Wonmin} and to outperform classical computers~\cite{Bartlett02,Bartlett,Ohliger,Mari12,Veitch}.  These applications have kindled an interest in the idea of Wigner function negativity as a resource~\cite{Mari11}.}

Until now, known protocols that Gaussify and distill entanglement have the feature of being recursive. To execute these protocols to greater depth requires greater memory storage requirements.  The quantum states are combined via a tree like process of pairwise distillation, with each branch demanding additional memory.  In the finite dimensional setting, entanglement pumping protocols~\cite{DB02a,CTSL01a,DCJTMJZHL01a,JTSL02a,JTSL02a,CB01a,C01a,Campbell10,Benj12,Fujii12} offer the option of compressing the spatial memory requirement, even down to 3-4 qubits per location, at the cost of reduced efficiency and increased temporal overheads.  Recently, a continuous variable analog of entanglement pumping, the compact distillery scheme, has been proposed~\cite{Datta12}.   This scheme requires storage of only 2 modes per location at any moment in time.   However, this pumping protocol is not a direct analog of the Gaussification protocols.  In particular, the compact distillery does not Gaussify and allows only a modest increase in entanglement.

Here, we extend and further develop the techniques of Ref.\ \cite{Campbell12} where the class of Gaussification protocols was vastly broadened and shown to work in virtue of quantum central limit theorems.   This work broadens the class of Gaussifer protocols, and in doing so introduces the concept of a pumping-Gaussifier that only requires 2 modes of memory per location.  Unlike the compact distillery scheme, our pumping protocol still Gaussifies and is capable of the same large increases of entanglement possible with the recursive-Gaussifier.  Surprisingly, the pumping-Gaussifer outputs the same final state as the more well known recursive-Gaussifiers.   This makes pumping-Gaussifers extremely promising protocols that are especially attractive for experiments with only a small number of modes.  We also comment on implications of our findings to devising novel schemes for long distance quantum communication via quantum repeater networks. Despite considerable research on CV entanglement distillation, surprisingly these techniques have not previously been explicitly applied to design of quantum repeaters.  Indeed, here, we provide the first concrete evidence in the CV context that using quantum repeaters can achieve greater distances of communication than direct transmission.

On a technical level, the approach taken here is complementary to, but subtly distinct from, our earlier results \cite{Campbell12}. In particular, compared to these earlier results, the relationship between quantum central limit theorems and Gaussification protocols requires a smaller and simpler set of assumptions required of the physical system. Center stage is taken by a class of non-commutative central limit theorems, which are general enough to capture all of the aforementioned situations of state manipulation, including post-selecting measurements. The requirements for a quantum central limit theorem to be valid will be highlighted and discussed in great detail.  We remark that these techniques are closely related to those used to prove the extremality principle~\cite{Wolf06};  which asserts that for entanglement measures satisfying very specific properties~\footnote{Note that these properties fail for some entanglement measures.   For instance, not only do they fail for log-negativity, but it is known that some non-Gaussian states do have less log-negativity than the equivalent Gaussian state (e.g. see Ref.~\cite{Wolf06}).}, Gaussian states have the least entanglement of all states with the same second moments.

\section{Continuous variable systems and phase space}

Here, we introduce our notation and briefly introduce some phase space concepts used throughout.  For more details see Refs.\ \cite{BR01b,Eisert03,Kok07}.  For a single mode of a continuous variable (herein CV) system, two important observables are
\begin{eqnarray}
	\hat{X} & = &  (\hat{a} + \hat{a}^{\dagger})/\sqrt{2} , \\
	\hat{P} & = &  i(\hat{a}^{\dagger} - \hat{a})/\sqrt{2} ,
\end{eqnarray}
that are analogs of position and momentum in simple harmonic oscillators, with $\hat{a}$ and $\hat{a}^{\dagger}$ 
being the photonic annihilation and creation operators.  For $m$ optical modes, the set of $2m$ quadrature operators is denoted as a vector of operators
\begin{equation}
	 \vec{\hat{Q}} =  (\hat{Q}_{1}, \hat{Q}_{2},\dots, \hat{Q}_{2m-1}, \hat{Q}_{2m}) = 
	 ( \hat{X}_{1}, \hat{P}_{1},,\dots, \hat{X}_{m}, \hat{P}_{m}).
\end{equation}
For a quantum state $\rho$, the expectation values of these quadratures are denoted by a set of $2m$ real numbers
\begin{equation}
	[ \vec{d}_{\rho} ]_{k} = \tr ( \hat{Q}_{k} \rho ),
\end{equation}
which are called the first moments of $\rho$.   Typically, we are interested in states with zero first moments, so $\vec{d}_{\rho}=0$.  The second moments, akin to variances, are captured by the covariance matrix
\begin{equation}
	[\Gamma_{\rho}]_{j,k} = 2 \Re \{ \tr [ ( \hat{Q}_{j} - [\vec{d}_{\rho}]_{j}  )( \hat{Q}_{k} - [\vec{d}_{\rho}]_{k}  ) \rho] \},
\end{equation}
which for states with zero first moments simplifies to
\begin{equation}
	[\Gamma_{\rho}]_{j,k} =  \tr [ ( \hat{Q}_{j} \hat{Q}_{k} + \hat{Q}_{k} \hat{Q}_{j}  ) \rho].
\end{equation}
It is easy to verify that, for physical states, the covariance matrix is real and symmetric.  

The first and second moments only partially describe the quantum state, but a complete description can be achieved by using one of a plethora of phase space representations.  In particular we make use of characteristic functions
$\chi_{\rho}:\R^{2m}\rightarrow\C$ such that
\begin{equation}
	\chi_{\rho} ( \vec{r} ) = \tr [ D ( \vec{r} ) \rho ], 
\end{equation}
where $D_{\vec{r}}$ is the unitary displacement or Weyl operator
\begin{equation}
	D_{\vec{r}} = \exp ( i \vec{r}.\hat{\vec{Q}}) =   \exp \left( i  \sum_{j } r_{j} \hat{Q}_{j} \right) .
\end{equation}
We say a state is Gaussian if and only if its characteristic function has a Gaussian shape, which entails
\begin{equation}
	\chi_{\rho} ( \vec{r} ) = \exp (  i \vec{r}.\vec{d}_{\rho} - \vec{r}^{T} \Gamma_{\rho} \vec{r}/4 ).
\end{equation}
Any state outside this set is said to be non-Gaussian.  Notable Gaussian states include the vacuum and the coherent states.  
The Wigner function, which is perhaps more widely known, is simply the Fourier transform of the characteristic function.  Since the Fourier transform maps the set of Gaussian functions to itself, the definition of Gaussian states is equivalent if stated in terms of Wigner functions.  For our purposes the characteristic function is the most useful choice of phase space representation.

Regarding dynamics, we say a unitary is Gaussian if it has the form $U = \exp ( i  H )$ where $H$ is Hermitian and quadratic in annihilation and creation operators.  The canonical example of a Gaussian measurement is a homodyne, or quadrature, measurement of an observable $\hat{Q}_{j}$.  More general Gaussian measurements can be related to quadrature measurements by use of Gaussian unitaries and ancillary Gaussian states.   For example, so called 8-port homodyne measurements project onto the coherent states and can be implemented by using two quadrature measurements and an ancillary mode in the vacuum state.

The most general kind of Gaussian operations are Gaussian channels (completely positive maps).  This class of physical operations are most naturally defined by using the Choi-Jamiolkowski (CJ) isomorphism \cite{Jam01a,ZyB01a} between quantum states and channels.  For a channel, $\mathcal{E}$, mapping $m$-mode quantum states to $m$-mode quantum states, the CJ-state is
\begin{equation}
	\Phi_{\mathcal{E}} = (\id \ox \mathcal{E}) \Phi  ,
\end{equation}
where $\Phi = \kb{\phi}{\phi}^{\otimes m}$ is a pure unnormalized operator with
\begin{equation}
	\ket{\phi} = \sum_{n=0}^{\infty} \ket{n, n} .
\end{equation}
Conversely, for all $\Phi_{\mathcal{E}}$ there exists a unique quantum channel $\mathcal{E}$ specified by the isomorphism, such that
\begin{equation}
	\mathcal{E}_{\Phi}(\rho) = \tr^{B} \left[ \Phi^{T_{B}}( \id \ox \rho)  \right] ,
\end{equation}
where $T_{B}$ is a partial transpose with respect to $B$.  When the Gaussian CP map acts on a Gaussian state, $\rho$, with covariance matrix, $\Gamma_{\rho}$, it has been shown~\cite{Eisert03, NoGoFu,Eisert03,Giedke02} that the output state $\rho'$ is also Gaussian with covariance matrix
\begin{equation}
	\Gamma_{\rho'} =   \gamma_{AA} - \gamma_{AB} (  \gamma_{BB} + \Gamma_{\rho})^{-1} \gamma^{T}_{AB} ,
\end{equation}
where 
\begin{equation}
\label{Eq_gamma}
	\gamma = \left( \begin{array}{cc} \gamma_{AA} & \gamma_{AB} \\  \gamma_{AB}^{T} & \gamma_{BB} \end{array} \right)  ,
\end{equation}
is the covariance matrix of $\Phi^{T_{B}}$ shown as a block matrix with respect to the partition between systems $A$ and $B$.  The expression for $\Gamma_{\rho'}$ takes the form of a Schur complement, which often arises in matrix problems and Gaussian integration \cite{HJ01b}. The partial transpose has a simple effect on covariance matrices, and so explicitly calculating the partial transposed state can be circumvented.  Partial transposition, in the Heisenberg picture, takes $\hat{P} \mapsto - \hat{P}$ for every momentum operator acting on system $B$.   Assume we know $\Phi_{\mathcal{E}} / \tr ( \Phi_{\mathcal{E}} )$ and its covariance matrix $\tilde{\gamma}$. It follows that the partial transposed state, $\Phi_{\mathcal{E}}^{T_{B}} / \tr ( \Phi_{\mathcal{E}} )$, has covariance matrix $\gamma = \Lambda \tilde{\gamma} \Lambda$ where $\Lambda = \id_{A} \oplus T_{B}$ and $T_{B}= \mathrm{diag}(1,-1,\dots, 1,-1)$.

\section{building blocks}
\label{secBBs}

This section introduces the basic building blocks of the protocols considered herein.  Each building block is specified by the following:  an operator, $\Pi$, called the filter; a value $R$ for the beam-splitter reflectivity; and a choice of two $m$-mode states that may be outputs from previous building blocks.  Throughout this article, any building blocks combined into a larger protocol will use the same filter $\Pi$, which must be an invertible operator proportional to a separable Gaussian state with zero first moments.  Such filters always, as shown in Ref.~\cite{Werner01}, have a decomposition of the form
\begin{equation}
\label{EqFilter}
	\Pi = \int P(\vec{r})  \Pi_{\vec{r}}
\end{equation}
where $P(\vec{r})$ is a classical, and Gaussian, probability distribution and
\begin{equation}
	  \Pi_{\vec{r}} = D_{\vec{r}} \kb{\psi}{\psi} D_{\vec{r}}^{\dagger} 
\end{equation}
for some pure separable Gaussian $\ket{\psi}$.   The set of operators  $\{  \Pi_{\vec{r}} \}$ specifies the POVM measurement to be used in the building block.   Recall, that 8-port homodyne measurements implement a similar POVM where $\ket{\psi}$ is the vacuum state, and so the desired POVM is always equivalent, up-to a local Gaussian unitary, to 8-port homodyne measurement.  The weighting $P(\vec{r})$ is a function of the measurement outcome, $\Pi_{\vec{r}}$, and dictates the postselection strategy used in the building block.   Another important special case is the one where $\Pi$ approximates the vacuum arbitrarily well, which is the situation considered in Ref.~\cite{Browne03,Eisert04}.

\begin{figure}
\includegraphics{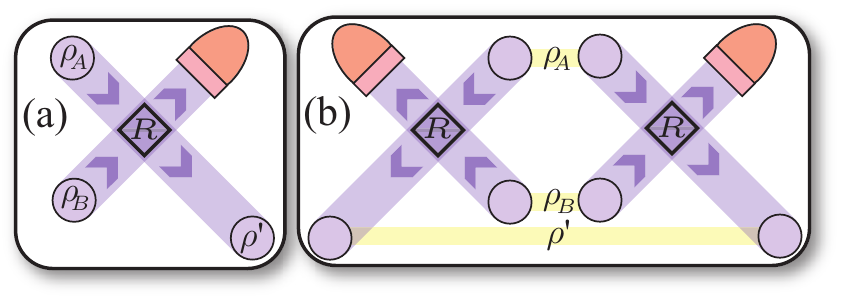}
\caption{An implementation of an individual building block with beam-splitter reflectivity $R$ for (a) single mode states and (b) two mode states.  Generalization to $m$-mode states is straightforward as each additional party performs the same local unitaries.}
\label{Fig_Blocks}
\end{figure}

Implementation of a building block is outlined in Fig.~\ref{Fig_Blocks} and is as following:
\begin{enumerate}
	\item Take two $m$-mode quantum states $\rho_{A}$ (modes $Aj$) and $\rho_{B}$ (modes $Bj$) ;
	\item Each of the $m$ parties mixes their two modes on a beam-splitter of reflectivity $R$;
	\item On each of the beam-splitters, take the output from the $B$-modes and locally implement the Gaussian measurement with local POVM elements $\{ \Pi_{\vec{r}} \}$
	\item  Given measurement outcome data $\vec{r}$, postselect declaring a success with probability $P(\vec{r})$;
	\item  Take the unmeasured $A$-modes and output from the building block.
\end{enumerate}
Here, we have labelled the $2m$-modes as $\{A1,,\dots, Am, B1,\dots, Bm\}$ and modes sharing the same numerical index share the same physical location. When successful, the building block outputs a state
\begin{equation}
	\rho' \propto \int P(\vec{r}) \tr_{B} [  U ( \rho_{A} \ox \rho_{B} ) U^{\dagger}  ( \id \otimes \Pi_{\vec{r}}  ) ] d \vec{r}.
\end{equation}
The unitary, $U$, represents the effect of the beam-splitters such that for all $j$
\begin{equation}
\label{Unitary}
	U^{\dagger } \hat{a}_{Aj} U = \sqrt{T} \hat a_{Aj} +  \sqrt{R} \hat a_{Bj} , 
\end{equation}
where $T=1-R$.  Taking the integral over measurement outcomes inside the partial trace and using Eq.~(\ref{EqFilter}) we have
\begin{equation}
	\rho'  \propto \tr_{B} [  U ( \rho_{A} \ox \rho_{B} ) U^{\dagger}  ( \id \otimes \Pi ) ] .
\end{equation}
Unfortunately,  the effect of this map can be difficult to analytically evaluate. The root of the technicalities are related to the 
fact that $U$ and $\id \otimes \Pi$ do not commute.
However, following the insights of Ref.\ \cite{Campbell12}, we know that by moving to phase space and working with a different object from $\rho'$ the effect of the map can be simplified. 
This is the key insight that renders the analysis feasible. In Ref.\  \cite{Campbell12} the characteristic function of the non-Hermitian object $\rho' \Pi$ was considered.   This work follows parallel reasoning but instead consider the Hermitian object $\Pi^{1/2} \rho' \Pi^{1/2}$ and its characteristic function. We make use of 
\begin{equation}
	\tau' = \frac{ P \rho' P }{  \tr ( P \rho'P ) }
\end{equation}
for the normalized and Hermitian filtered object, with
\begin{equation}
	P=\Pi^{1/2}.  
\end{equation}
This object is then
\begin{equation}
	\tau' \propto \tr_{B}[  ( P \ox \id ) U ( \rho_{A} \ox \rho_{B} ) U^{\dagger}  ( P \ox \Pi ) ].
\end{equation}
Splitting $\Pi = P P$ and using the cyclicity of the trace we have the more symmetric formula
\begin{equation}
	\tau' \propto \tr_{B} [ (P \ox P  )U ( \rho_{A} \ox \rho_{B} ) U^{\dagger}  ( P \ox P ) ].
\end{equation}
The next fact we employ is that for any Gaussian operator with zero-first moments, such as $P$, we have that
\begin{equation}
	U^{\dagger} (P \ox P) U   =  P \ox P .
\end{equation}
This equality is well known (see e.g. Ref.~\cite{Olivares11}), 
but for completeness we give a proof in App.\ \ref{App_Comm_Lemma}.   Hence we have
\begin{equation}
	\tau' \propto \tr_{B} [ U ( P\rho_{A}P \ox P\rho_{B}P ) U^{\dagger}  ].
\end{equation}
Again using the shortened notation, $\tau_{A} \propto P \rho_{A}P $  and $\tau_{B} \propto P \rho_{B}P $, gives
\begin{equation}
	\tau' \propto \tr_{B} [ U (\tau_{A} \ox \tau_{B} ) U^{\dagger}  ] .
\end{equation}
By choosing $\Pi$, and equivalently $P$, as proportional to a Gaussian state, we have been able to exploit the symmetry of the problem to reach a greatly simplified expression.  The characteristic function of this object is then
\begin{equation}
	\chi_{\tau' }(\vec{r}) \propto \tr [ (\id \otimes D_{\vec{r}}) U (\tau_{A} \ox \tau_{B} ) U^{\dagger} ].
\end{equation}
Conjugating $U^{\dagger}$ with the displacement operator gives
\begin{eqnarray}
	U^{\dagger}  (\id \ox D_{\vec{r}} )U & = &  U^{\dagger} \exp [ i  ( \id \ox \vec{r}.\vec{\hat{Q}} ) ] U , \nonumber \\ \nonumber 
  & = &  \exp [ i  \sqrt{T} ( \vec{r}.\vec{\hat{Q}} \ox \id) + i \sqrt{R} ( \id \ox \vec{r}.\vec{\hat{Q}} ) ] , \\ 
& = &	  	  D_{\sqrt{T} \vec{r}} \ox D_{\sqrt{R}\vec{r}} . 
\end{eqnarray}
Using this relation we deduce that
\begin{eqnarray}
	\chi_{\tau' }(\vec{r}) & \propto & \tr [ D_{\sqrt{T} \vec{r}} \tau_{A} \otimes D_{\sqrt{R} \vec{r}} \tau_{B}   ]. \\ \nonumber
				& \propto & \tr [ D_{\sqrt{T} \vec{r}} \tau_{A} ] \tr  [ D_{\sqrt{R} \vec{r}} \tau_{B}   ]. 
\end{eqnarray}
However, these factors are simply the characteristic functions for $\tau_{A}$ and $\tau_{B}$ but with a modified value of $\vec{r}$, so
\begin{eqnarray}
	\chi_{\tau' }(\vec{r}) & = &   \chi_{\tau_{A}}(\sqrt{T} \vec{r}) \chi_{\tau_{B}}(\sqrt{R} \vec{r}) .
\end{eqnarray}
We have shifted to equality, rather than proportionally, because the characteristic function of a unit trace object takes $\chi_{\tau}(0)=1$.  As promised, the effect of the protocol on the filtered $\tau$ objects is much more straightforward than for the actual density matrices.   Note that if we considered non-Hermitian objects $\sigma_{A,B} \propto \rho_{A,B} \Pi$ and output $\sigma' = \rho' \Pi$, we would have similarly arrived at
\begin{eqnarray}
\label{GeneralIter}
	\chi_{\sigma' }(\vec{r}) & = &   \chi_{\sigma_{A}}(\sqrt{T} \vec{r}) \chi_{\sigma_{B}}(\sqrt{R} \vec{r}) .
\end{eqnarray}
These results generalize those of Ref.\ \cite{Campbell12}, where the input states were taken to be identical and reflectivity set to be $50/50$ and only the non-Hermitian objects were considered.  Later in this article, we find that working with Hermitian objects proves to be the more elegant approach.

Before proceeding we remark on the assumption that $\Pi$, and hence all $P$, are invertible.  The assumption is required to ensure that $\tau'$ uniquely define $\rho'$.  All Gaussian operators, except projectors, are full rank and invertible so the assumption simply rules out projectors.  However, we wish for our general analysis to encompass previous protocols \cite{Browne03,Eisert04,Datta12} that prescribe projecting two modes onto the vacuum, where $\Pi=P=\kb{0,0}{0,0}$, which is clearly not invertible.  However, any realistic experiment will use detectors with some non-unit efficiency of photon detection.   Indeed, often efficiency is significantly less than unity.  Such inefficiencies can be modelled by placing a beam-splitter
ahead of the detector, and can be easily incorporated into our analysis.  This modification results in a realistic filter that is still Gaussian but no longer a projector.  As such, the assumption of invertible filters is always justified.

\section{Protocols}

\subsection{The recursive-Gaussifer}

The first class of protocols we review were originally introduced in Ref.\ \cite{Campbell12}, generalizing the proposals of Refs.\ \cite{Browne03,Eisert04}.  We refer to the protocols considered here as recursive-Gaussifers and the general structure is outlined in Fig.~(\ref{Fig_ComposingBlocks}a).  All building blocks of the recursive protocol use the same filter $\Pi$, and set $R=T=1/2$.   In the first round of the protocol many copies of a raw state $\rho_{1}$ are taken and are simultaneously used as inputs to building blocks, with $\rho_{A}=\rho_{B}=\rho_{1}$.  The successful outputs from these rounds are labelled $\rho_{2}$, and are used as the inputs into the building blocks for the next round.  On the $n^{\mathrm{th}}$ round, each building block takes two input states labelled $\rho_{2^{n}}$ and outputs $\rho_{2^{n+1}}$.  The subscript counts the number of raw copies so far consumed.  Denoting  $\tau_{2^{n}}\propto P \rho_{2^{n}}P$ and applying Eq.~(\ref{GeneralIter}) we find that
\begin{equation}
	\chi_{\tau_{2^{n+1}}} (\vec{r}) = \chi_{\tau_{2^{n}}} \left( \frac{\vec{r}}{\sqrt{2}}\right)^{2},
\end{equation}
which is easier to represent in terms of $N=2^{n}$ so
\begin{equation}
	\chi_{\tau_{2N}}  (\vec{r}) = \chi_{\tau_{N}} \left( \frac{\vec{r}}{\sqrt{2}}\right)^{2}.
\end{equation}
In terms of $\tau_{1}$ we have 
\begin{equation}
	\chi_{\tau_{N}}  (\vec{r}) = \chi_{\tau_{1}} \left( \frac{\vec{r}}{\sqrt{N}}\right)^{N},
\end{equation}
To reach $n$ rounds, assuming every building blocks succeeds, we must have a memory capable of storing $N=2^{n}$ copies of $\rho_{1}$ simultaneously.  The exponential increase in memory is required because we have assumed simultaneous execution of all building blocks within a round.  However, relaxing the simultaneity requirement and using a smart ordering --- for instance as in Fig.~(\ref{Fig_ComposingBlocks}b) --- the recursive protocol can implement $n$ rounds with a storage capacity of $n+1$ modes per location, albeit at the cost of increasing the number of time steps.   A growing quantum memory seems unavoidable, but we will soon see how it can be circumvented.
The sequence of characteristic functions 
\begin{equation}
	\{ \chi_{\tau_{1}} , \chi_{\tau_{2}},  \chi_{\tau_{4}} ,  \chi_{\tau_{8}}, \dots \} 
\end{equation}
 is known to evolve toward a Gaussian with unchanged second moments by virtue of a central limit theorem.  We will later review central limit theorems, providing extensions to make more direct statements about the physical state. 

\subsection{The pumping-Gaussifier}

\begin{figure}
\includegraphics{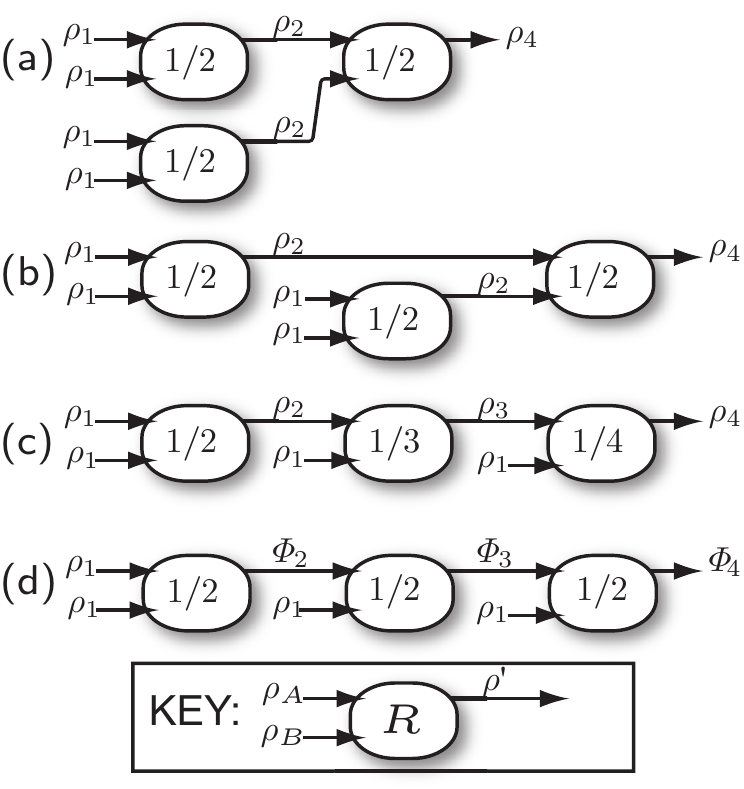}
 \caption{Different protocols combining building blocks in ways.  All building blocks use the same filter, $\Pi$, and are labelled with their beam-splitter reflectivity $R$. (a) the recursive-Gaussifier protocol; (b) the temporally-reordered recursive-Gaussifier protocol; (c) the pumping-Gaussifier protocol; and (d) the compact distillery protocol.  The key show how the building block labels compare with the variables used in  Sec~\ref{secBBs}.}
 \label{Fig_ComposingBlocks}
 \end{figure}

Our next class of protocols are entirely novel.   We propose protocols that use a fixed initial state to repeatedly pump a target state, surprisingly resulting in the same output as an analogous recursive protocol.  The building blocks that compose the pumping-Gaussifier use two distinct input states in later rounds and also weaken the beam-splitter reflectivity with the number of steps. On the $N^{\mathrm{th}}$ step, we take a copy of $\rho_{N}$ and a raw initial state $\rho_{1}$ and mix on a beam-splitter of reflectivity $R_{N}=1/(N+1)$ as shown in Fig.~(\ref{Fig_ComposingBlocks}c).  The output is labelled $\rho_{N+1}$ and in the phase space picture we have the iterative formula
\begin{equation}
	\chi_{\tau_{N+1}}(\vec{r}) = \chi_{\tau_{N}} \left( \frac{\sqrt{N}}{\sqrt{N+1}}  \vec{r} \right) \chi_{\tau_{1}}\left( \frac{1}{\sqrt{N+1}} \vec{r} \right) ,
\end{equation}
We can verify that 
\begin{equation}
	\chi_{\tau_{N}}(\vec{r}) =   \chi_{\tau_{1}}\left( \frac{\vec{r}}{\sqrt{N}} \right)^{N} ,
\end{equation}
satisfies the iterative formula because
\begin{eqnarray}  \nonumber
	\chi_{\tau_{N+1}}(\vec{r}) &=& \chi_{\tau_{1}}     \left( \frac{\sqrt{N}}{\sqrt{N+1}} \frac{\vec{r}}{\sqrt{N}} \right)^{N}       \chi_{\tau_{1}}\left( \frac{\vec{r}}{\sqrt{N+1}} \right) , \\ \nonumber
	& = &\chi_{\tau_{1}} \left( \frac{\vec{r}}{\sqrt{N+1}}  \right)^{N}  \chi_{\tau_{1}} \left( \frac{\vec{r}}{\sqrt{N+1}} \right) , \\ 
	& = &\chi_{\tau_{1}} \left( \frac{\vec{r}}{\sqrt{N+1}}  \right)^{N+1}.
\end{eqnarray}
The neat cancellation of $\sqrt{N}/\sqrt{N}$ only occurs because of our exact choice of beam-splitter reflectivity.  After the $N^{\mathrm{th}}$ step,  the characteristic function matches that of the recursive-Gaussifier implemented to depth $n=\log_{2}(N)$.  Furthermore, for successful implementations both protocols consume the same number of raw copies to achieve the same output.  However, in the pumping protocol we also have the option of terminating after a number of steps not of the form $N=2^{n}$.

\subsection{The compact distillery}

The compact distillery (CD) protocol~\cite{Datta12} also repeatedly pumps with the same initial state, but it keeps a constant beam-splitter reflectivity of $R=1/2$ as outlined in  Fig.~(\ref{Fig_ComposingBlocks}d). The CD protocol is known to provide a very different evolution from both our Gaussifier protocols.  To highlight that it produces different states from the Gaussifiers, we label the output of the $N^{\mathrm{th}}$ step as $\Phi_{N+1}$ and equate $\Phi_{1}=\rho_{1}$ for the raw resource. Denoting $\phi_{N} \propto P \Phi_{N} P$ we have the iterative relation
\begin{equation}
	\chi_{\phi_{N+1}}(\vec{r})  = \chi_{\phi_{N}} \left( \frac{\vec{r}}{\sqrt{2}} \right)  \chi_{\phi_{1}}  \left( \frac{\vec{r}}{\sqrt{2}} \right) .
\end{equation}
We can immediately deduce properties of $\phi_{N+1}$ from those of the initial operator $\phi_{1}$.  For instance, if the characteristic function $ \chi_{\phi_{1}} $ is zero at point $\vec{r}_{0}$ then the characteristic function $ \chi_{\phi_{N+1}} $ is zero at $\sqrt{2}\vec{r}_{0}$ for all $N$.  Hence, the characteristic function $\chi_{\phi_{N+1}}$ will not have a Gaussian shape and consequently the corresponding physical state, $\Phi_{N+1}$, will also be non-Gaussian.  If there exists a limiting characteristic function, $\chi_{\phi_{\infty}}$, the same argument applies and so non-Gaussianity would persist even in the asymptotic limit of many iterations.   Indeed, all the examples considered in Ref.\ \cite{Datta12} found that the protocol converges toward non-Gaussian states.  Our phase space techniques provide a clear explanation of \textit{why} non-Gaussianity persists in the compact distillery.  This illustrates the merit of the phase space perspective, even for examining protocols that do not Gaussify.  

The CD protocol was proposed as an alternative to recursive-Gaussifiers to reduce the required quantum memory and bring protocols closer to experimental feasibility.  However, we have seen that our pumping-Gaussifier can also operate under these stringent memory constraints.  We must then consider other figures of merit to compare these protocols.  The authors of Ref.\ \cite{Datta12} showed that, when feed with weakly entangled photon subtracted states, a few rounds of the CD achieves a similar entanglement increase as a few rounds of the Gaussifier.  However, the maximum achievable entanglement of the Gaussifer proved to be much higher, and so after only 3-4 rounds the advantage of the pumping-Gaussifier can be significant.  Of course, whether we desire the output state to be non-Gaussian or Gaussian depends on the context and what quantum information protocol the resource is subsequently used for.

\section{Central Limit Theorems}

\subsection{Characteristic function convergence}

Central limit theorems are results that tell us when a sequence of characteristic functions approaches a Gaussian function and in what way they converge.  Throughout we are interested in sequences of characteristic functions output by the recursive and pumping-Gaussifers.
\begin{defin}[Central limit sequence]
We say a sequence of Hermitian positive operators $\{ \tau_{N} \}$ and associated characteristic functions  $\{ \chi_{\tau_{N}}\}$ is a central limit sequence if 
 \begin{equation}
  	\chi_{\tau_{N}}  (\vec{r}) = \chi_{\tau_{1}} \left( \frac{\vec{r}}{\sqrt{N}} \right)^{N} ,
 \end{equation}
where $\chi_{\tau_{1}}$ has zero first moments and $\Gamma_{\tau}$ second moments.
 \end{defin}
For such a sequence, if $\tau_{1}$ is Hermitian and positive then the results of Refs.\ \cite{Hudson71,Wolf06,Campbell12} govern its limiting behaviour.  More generally, if $\tau_{1}$ is non-Hermitian then recent results~\cite{Campbell12} give conditions under which it approaches a Gaussian.  These latter techniques were used to demonstrate Gaussification of physical systems by considering $\tau_{1} \propto \rho_{1}\Pi$.  Here we consider the Hermitian $\tau_{1}  \propto P \rho_{1} P$ for which the convergence properties are simpler to state.  

 \begin{theorem}[General quantum central limit theorem]
 \label{CLTchar}
Consider a central limit sequence $\{ \chi_{\tau_{N}}\}$. For any finite radius $r_{0}$ and any accuracy  $\epsilon>0$, there exists an $N_{\epsilon}$ such that for all $N\geq N_{\epsilon}$ and all $|\vec{r}|\leq r_{0}$ we have 
 \begin{equation}
	| \chi_{\tau_{N}}(\vec{r}) - \chi_{\tau_{\infty}}(\vec{r}) | < \epsilon.
\end{equation}
where $\chi_{\tau_{\infty}}(\vec{r})$ is a Gaussian with covariance matrix $\Gamma_{\tau_{1}}$.
\end{theorem}
The theorem can be proven by taking a cross section of the characteristic function for a unit direction $\vec{r}$, such that 
\begin{equation}
	f_{N}(t)=\chi_{\tau_{N}}(t \vec{r}) , 
\end{equation}
and proving convergence to a Gaussian function in phase space
for all such cross sections.  Each cross section is equivalent to a characteristic function for a classical probability distribution.  We may proceed by following one of the numerous classical proofs, such as Ref.\ \cite{Moran}.   Central limit theorems are fundamental to our method and so for completeness we will provide a proof here. 

From the definition of a characteristic function, it follows that it can be expanded as
\begin{equation}
	f_{1}(t) = 1 -\frac{ t^{2}}{2} \nu  + C(t^{2})
\end{equation}
where $\nu$ is the second moment in direction $\vec{r}$, such that
\begin{equation}
	\nu = 2 \tr [   (\vec{r}.\vec{\hat{Q}})^2 \rho ] ,
\end{equation}
and the higher order terms $C(x^{2})$ can be shown~\cite{Moran,Campbell12} to satisfy $C(x^{2})/x^{2} \rightarrow 0$ as $x \rightarrow 0$.  Hence, the $N^{\mathrm{th}}$ function in the sequence is
\begin{equation}
	f_{N}(t) = \left(1 -\frac{ t^{2}}{2N} \nu  + C ( t^{2}/ N  ) \right)^{N} .
\end{equation}
We wish to compare this with $\exp(-t^{2} \nu)$, and so the difference of these quantities is
\begin{equation}
	\delta_{N}(t) = |	f_{N}(t)  - \exp(-t^{2} \nu) | .
\end{equation}
We can approximate $\exp(-t^{2} \nu)$ with some $(1-t^{2} \nu/N)^{N}$ to any accuracy $\epsilon/2>0$, such that there exists an $N'_{\epsilon}$ and for $N>N'_{\epsilon}$ we have
\begin{eqnarray*}
	\delta_{N}(t) & \leq & \left|  \left(1- \frac{t^{2}\nu}{N} + C \left( \frac{t^{2}}{N} \right) \right)^{N}- \left(1- \frac{ \nu }{N} \right)^{N} \right| + \frac{\epsilon}{2} .
\end{eqnarray*}
Next, we use that for any complex numbers, $a$ and $b$, with $|a|\leq 1$ and $|b|\leq 1$ we know (see App.\ \ref{miniIdentity}) that 
$|a^{N}-b^{N}|\leq N |a-b|$ and applying this yields
\begin{eqnarray}
	\delta_{N}(t) & \leq & N \left| \left(1- \frac{t^{2}\nu}{N} + C \left( \frac{t^{2}}{N} \right)\right) +\left(1- \frac{\nu}{N}\right)  \right| + 
	\frac{\epsilon}{2} , \nonumber \\ 
	 & = &  N | C \left( t^{2}/ N \right) | +\frac{\epsilon}{2} , \nonumber \\ 
	 &  = & t^{2} | C \left( x^{2} \right) / x^{2} | + \frac{\epsilon}{2} ,
\end{eqnarray}
where $x^{2}=t^{2}/N$.  For constant $t$, we can decrease $x$ to any desired value by increasing $N$.  Since $C \left( x^{2} \right) / x^{2}$ vanishes in this limit, for any desired $\eta=\epsilon / 2 t^{2} > 0$ we can find a $N''_{\epsilon}$ such that for all $N>N''_{\epsilon}$ we have $|C \left( x^{2} \right) / x^{2} | \leq \eta$.  Hence, we have 
\begin{equation}
	\delta_{N}(t) \leq t^{2} \eta + \epsilon/2 = \epsilon .
\end{equation}
This final result holds for $N> \max ( N'_{\epsilon}, N''_{\epsilon})=N_{\epsilon}$.   The above argument tells us how individual points evolve in $N$, but the result can be strengthen further for all points within a ball of finite radius $r_{0}$.  This extension to finite regions of phase space is outlined in App~\ref{APPuniform}.  This result is stronger as the same error bound uniformly holds across a whole region simultaneously.  The region has a finite area and extensions of this result to the whole of phase space do not hold. Indeed, central limit theorems are aptly named as they dictate the limiting behaviour around the origin of phase space but \textit{not} into the tails (see also the similar discussion related to non-commutative central limit theorems applied to grasping quantum many-body dynamics \cite{QCLT,QCLTOld}).

\subsection{Convergence of moments}

Next, we present a second aspect of central limit theorems, which we use later, that quantifies the evolution of higher moments.  We begin by generalizing the idea of a quadrature.  Typically, quadratures are thought of as single mode position or momentum operators, but we take quadratures to include all linear combinations of such operators, such that
\begin{equation}
\label{quadrature}
	H = \sum_{j} r_{j} \hat{Q}_{j} ,
\end{equation}
is always a quadrature.  The $k^{\mathrm{th}}$ moment of such an operator, assuming first moments are zero, is the expectation value of $H^{k}$.  More generally, we say an operator is a $k^{\mathrm{th}}$ moment if it is a product of $k$, potentially distinct, quadratures, such that
\begin{equation}
	H^{(k)} = \prod_{j=1}^{k} H_{j} ,
\end{equation}
where each $H_{j}$ is linear in quadrature operators as in Eq.~(\ref{quadrature}).  Another result known as a central limit theorem is the following.
 \begin{theorem}[Convergence of moments]
 \label{momentsConverge}
For any central limit sequence $\{ \tau_{N} \}$ and any $k^{\mathrm{th}}$ moment, $H^{(k)}$, in the large $N$ limit
 \begin{equation}
 	| \tr ( H^{(k)}  \tau_{N} ) -  \tr ( H^{(k)}  \tau_{\infty} ) | \rightarrow 0 .
 \end{equation}
\end{theorem}
A simplified proof of this result is presented in App.~\ref{APPmoments}, but more involved proofs of more general results can be found in Refs.\ \cite{Giri78,Petz90}.  The theorem can be easily extended to finite linear sums of moments as follows.
\begin{corollary}[Finite sums of moments]
Consider an operator, $H$, which is a sum of finitely many terms, each a $k^{\mathrm{th}}$ moment.   The sequence of operators $\tau_{N}$ for increasing $N$ obeys
 \begin{equation}
 	| \tr ( H  \tau_{N} )- \tr (  H  \tau_{\infty} ) | \rightarrow 0 .
 \end{equation}
\end{corollary}

\subsection{Matrix element convergence}

The above theorems tell us about the evolution of the characteristic functions and moments but what can be said on the level of the density matrices $\tau_{N}$?  We have the following:
 \begin{theorem}[Pointwise convergence]
 \label{pointwiseMatrix}
 Consider a central limit sequence $\{  \tau_{N} \}$ and a pair of pure states $\{ \ket{\psi_{k}} , \ket{\psi_{j}} \}$,  in the limit of large $N$ 
 \begin{equation}
	| \bra{\psi_{k}} \tau_{N} \ket{\psi_{j}} - 	\bra{\psi_{k}} \tau_{\infty} \ket{\psi_{j}} |  \rightarrow 0.
\end{equation}
\end{theorem}
This tells us that individual matrix elements converge toward a fixed value and we give a proof in App.~\ref{APPMatrixElements}.  This result informs us of the evolution of the filtered object $\tau_{N} = P \rho_{N} P / \tr ( P \rho_{N} P )$.  However, we really want to know about the physical state $\rho_{N}$, and this is the problem we turn to in the next section.

\section{Convergence of physical state}

Knowing the filtered object obeys a central limit theorem, we can draw conclusions on the evolution of the actual physical state.  Recall that earlier we demanded, without loss of generality, that $P$ was an invertible matrix.  This assumption allows us to conclude that there exists a unique operator, 
\begin{equation}
	\rho_{N} \propto P^{-1} \tau_{N} P^{-1} .
\end{equation}
Concerning these states we shall show the following.
\begin{theorem}[State convergence]
\label{ThmPhysical}
Consider a central limit sequence $\{ \tau_{N} \}$ with limiting Gaussian operator $\tau_{\infty}$ and covariance matrix $\Gamma_{\tau}$. Denote $\gamma$ as the covariance matrix of the CJ-state --- see Eq.~(\ref{Eq_gamma}) --- isomorphic to the channel $\mathcal{P}$, such that $\mathcal{P}(\rho)=P \rho P$ for some Gaussian $P$.  If the covariance matrix
\begin{equation}
\label{Eq_GammaRelation}
	\Gamma_{\rho_{\infty}} = \gamma_{AB}^T(  \gamma_{AA} - \Gamma_{\tau_{\infty}})^{-1}  \gamma_{AB} - \gamma_{BB}
\end{equation}
exists and is physical, then $\rho_{\infty} \propto P^{-1} \tau_{\infty} P^{-1} $ exists and is a Gaussian state with covariance matrix $\Gamma_{\rho_{\infty}}$.  Furthermore, if  $\{ \ket{\psi_{k}} , \ket{\psi_{j}} \}$ are eigenvectors of $P$, then the sequence $\{ \rho_{N} \}$ in the large $N$ limit satisfies
\begin{equation}
	\left| \frac{ \bra{\psi_{k}}\rho_{N} \ket{\psi_{j}} }{\tr (P \rho_{N} P ) } - \frac{\bra{\psi_{k}} \rho_{\infty} \ket{\psi_{j}}}{\tr ( P \rho_{\infty} P ) }  \right| \rightarrow 0 .
\end{equation}
\end{theorem}
Above we define a limiting physical state and show a weak form of convergence of the density matrix elements up-to a normalisation factor.  It is worth noting that most existing results in the literature only go this far, though we will be interested in going further.
\begin{corollary}[Fidelity convergence]
\label{cor_fidcon}
In addition to Thm.\ \ref{ThmPhysical}, if also in the large $N$ limit we have $\tr (P \rho_{N} P ) \rightarrow \tr (P \rho_{\infty} P )$ then also
\begin{equation}
	F( \rho_{N} , \rho_{\infty}) \rightarrow 1 ,
\end{equation}
where $F$ is the fidelity between its arguments.
\end{corollary}
Let us prove this straightforward corollary. If $\tr (P \rho_{N} P ) $ converges to $\tr (P \rho_{\infty} P ) $, then we have that for increasing $N$
\begin{equation}
	| \bra{\psi_{k}}\rho_{N} \ket{\psi_{j}}  -  \bra{\psi_{k}} \rho_{\infty} \ket{\psi_{j}}  | \rightarrow 0.
\end{equation}
Furthermore, it is well-known that for physical states element-wise convergence of the density matrix entails converge in terms of fidelity and other measures of similarity such as trace norm distance \cite{Hudson71}.   However, the corollary rests upon additional key assumption that is the focus of the next section.

To prove our state convergence theorem we first find $\Gamma_{\tau_{\infty}}$ in terms of $\Gamma_{\rho_{\infty}}$, under the assumption that $\rho_{\infty}$ is Gaussian.  Since $P$ is invertible, there exists a unique physical state, defined by $P\rho_{\infty}P \propto \tau_{\infty} $.  In light of this uniqueness, the Gaussianity of $\rho_{\infty}$ is assured provided that a Gaussian solution to $P\rho_{\infty}P \propto \tau_{\infty} $ exists.
The operators are related by a CP-map, $A\mapsto \mathcal{P}(A)=P A P$
with Gaussian $P$, and so we can apply the results of Refs.\ \cite{Eisert03, NoGoFu,Eisert03,Giedke02} on Gaussian channels and the CJ isomorphism (reviewed earlier).  This tells us that for channel $\mathcal{P}$ with Gaussian CJ state acting on a Gaussian input state, the covariance matrices are related such that
\begin{equation}
	\Gamma_{\tau_{\infty}} = \gamma_{AA} - \gamma_{AB} (  \gamma_{BB} + \Gamma_{\rho_{\infty}})^{-1} \gamma^{T}_{AB},
\end{equation}
where $\gamma$ is as defined in Eq.~(\ref{Eq_gamma}).  To reach Eq.~(\ref{Eq_GammaRelation}) we simply rearrange the above expression for $ \Gamma_{\rho_{\infty}}$.  

Furthermore, denoting $\{ \ket{\psi_{j}}\}$ as the eigenvectors of $P$ with eigenvalue $\lambda_{j}$, we can apply Thm.~\ref{pointwiseMatrix} with respect to $\{ \ket{\psi_{j}} ,\ket{\psi_{k}} \}$.  Consequently, for large enough $N$
\begin{equation}
\label{weakConvergence}
	\left| \frac{  \lambda_{j} \lambda_{k}   \bra{\psi_{k}}\rho_{N} \ket{\psi_{j}} }{\tr (P \rho_{N} P ) } - 	\frac{  \lambda_{j} \lambda_{k} \bra{\psi_{k}} \rho_{\infty} \ket{\psi_{j}}}{\tr ( P \rho_{\infty} P ) }  \right| \rightarrow 0.
\end{equation}
After cancelling the $\lambda_{j} \lambda_{k}$ factors we have proven Thm.~\ref{ThmPhysical}.

\subsection{Convergence in fidelity}

In the previous section we made very general, but weak, predictions on the evolution of the physical state.  In order to deduce stronger conclusions, as captured by Cor. (\ref{cor_fidcon}), we need that $\tr ( P  \rho_{N} P ) $ converges to the value $\tr ( P  \rho_{\infty} P ) $.  Whether our protocols work correctly rests on the validity of this assumption.   The assumption appears fairly innocuous but is actually quite subtle, and, surprisingly, instances exist where it fails.  We remedy the neglect of this important assumption.

Some sufficient conditions have been found for this assumption~\cite{Campbell12}.  We strengthen these results, providing the basis for studies in subsequent sections.  Our result makes use of the idea of a reference state that we first define.

\begin{defin}[Reference state]
Consider an operator $\tau$ and a Gaussian filter $\Pi \propto \exp ( - \sum_{j} \beta_{j} \hat{b}_{j}^{\dagger} \hat{b}_{j} )$ where $\hat{b}_{j}=V \hat{a}_{j}V^{\dagger}$ for some Gaussian unitary $V$.  If $\tau_{\mathrm{ref}}$ is a Gaussian state, we write $\tau \leq_{\Pi} \tau_{\mathrm{ref}}$ if both of the following are satisfied:  
\begin{enumerate}
	\item[(i)] $| \tr ( H^{(k)} \tau ) | \leq  | \tr ( H^{(k)} \tau_{\mathrm{ref}} ) |$;
	\item[(ii)] $| \tr ( H^{(k)} \tau_{\mathrm{ref}} ) |  =  \tr ( H^{(k)} \tau_{\mathrm{ref}} )  $;
\end{enumerate}
for all moments $H^{(k)}$ composed of finite products of $\{ \hat{b}_{j}^{\dagger},  \hat{b}_{j} \}$.  When $\tau \leq_{\Pi} \tau_{\mathrm{ref}}$ we say $\tau_{\mathrm{ref}}$ is a reference state for $\tau$ w.r.t. $\Pi$.
\end{defin}
The concept is especially useful when considering central limit sequences because of the following.
\begin{lem}[Persistence of reference state]
\label{ThermalLem}
Consider a central limit sequence $\{ \tau_{N} \}$ and a Gaussian filter $\Pi$.   If there exists a $\tau_{j} \in \{ \tau_{N} \} $ and Gaussian $\tau_{\mathrm{ref}}$ such that $\tau_{j}  \leq_{\Pi }\tau_{\mathrm{ref}}$, then for all $N \geq j$ we have $\tau_{N}  \leq_{\Pi }\tau_{\mathrm{ref}}$.
\end{lem}
That the reference state remains good for all $N$ can be proven iteratively.  For any $k^{\mathrm{th}}$ moment, 
\begin{equation}
	\tr ( H^{(k)} \tau_{N+1}) = \tr [ U^{\dagger} (\id \otimes H^{(k)}) U ( \tau_{N} \otimes \tau_{1} ) ]  .
\end{equation}
The conjugation of $ H^{(k)}$ by $U$ gives a sum of $2^{k}$ terms, each a product of 
$\{ \hat{b}_{j}^{\dagger},  \hat{b}_{j} \}$ operators.  
We label each term by $x$, with it having the form $H^{(k-j_{x})}_{x} \otimes H^{(j_{x})}_{x}$ for some integer $j_{x}$ that depends on $x$.  In particular, for every $j$ the binomial "$k$ choose $j$" counts the multiplicity of $x$ values for which $j_{x}=j$.  In this notation 
\begin{eqnarray*}
	\tr ( H^{(k)} \tau_{N+1})  &= &  \sum_{x} C_{x} \tr [  (H^{(k-j_{x})}_{x} \otimes H^{j_{x}}_{x})  ( \tau_{N} \otimes \tau_{1} ) ] , \\ \nonumber
	& = & \sum_{x} C_{x} \tr (  H^{(k-j_{x})}_{x} \tau_{N} )  \tr (  H^{(j_{x})}_{x} \tau_{1} ) ,
\end{eqnarray*}
where $C_{x}= T_{N}^{(k-j_{x})/2} R_{N}^{j_{x}/2}$.  Assuming that the properties of reference states hold for $\tau_{N} $, we have for $\tau_{N} $ that
\begin{eqnarray*} \nonumber
	|\tr ( H^{(k)} \tau_{N+1})| & \leq & \sum_{x} C_{x} | \tr (  H^{(k-j_{x})}_{x} \tau_{N} ) |.| \tr (  H^{(j_{x})}_{x} \tau_{1} ) | , \\ \nonumber
	& \leq & \sum_{x} C_{x}  \tr (  H^{(k-j_{x})}_{x} \tau_{\mathrm{ref}} )  \tr (  H^{(j_{x})}_{x} \tau_{\mathrm{ref}} ) . 
\end{eqnarray*}
Next we recall that Gaussian states are invariant under the beam-splitter unitary, $U (\tau_{\mathrm{ref}} \otimes \tau_{\mathrm{ref}})U^{\dagger}=\tau_{\mathrm{ref}} \otimes \tau_{\mathrm{ref}}$, as was shown in App.\ \ref{App_Comm_Lemma}.   Being invariant under beam-splitters, Gaussian states must also be fixed points of the protocol and since $\tau_{\mathrm{ref}}$ is Gaussian we infer
\begin{equation}
     \tr ( H^{(k)} \tau_{\mathrm{ref}})  =  \sum_{x} C_{x}  \tr (  H^{(k-j_{x})}_{x} \tau_{\mathrm{ref}} )  \tr (  H^{(j_{x})}_{x} \tau_{\mathrm{ref}} ) .
\end{equation}
Using this invariance and applying it to the problem at hand we conclude
\begin{equation}
	|\tr ( H^{(k)} \tau_{N+1})|  \leq  \tr ( H^{(k)} \tau_{\mathrm{ref}}) .
\end{equation}
This proves, as claimed earlier, that when a reference state has the desired properties with respect to some $\tau_{j}$, it automatically follows for all $\tau_{N \geq j}$.  The concept of a reference state is fundamental to the following result.
\begin{theorem}[Convergence in fidelity]
\label{Strong}
Consider a central limit sequence $\{ \tau_{N}\}$ and filter $\Pi$.  If there exists a $\tau_{j} \in \{ \tau_{N}\}$ and Gaussian $\tau_{\mathrm{ref}}$ such that $\tau_{j} \leq_{\Pi} \tau_{\mathrm{ref}}$ and $\tr ( \Pi^{-1} \tau_{\mathrm{ref}} ) < \infty $, then
\begin{equation}
	\tr ( \Pi \rho_{N}  ) \rightarrow \tr ( \Pi \rho_{\infty}  ) 	,
\end{equation}
where $\rho_{N} = P^{-1} \tau_{N} P^{-1} / \tr ( P^{-1} \tau_{N} P^{-1}  )$.  Furthermore, as $N$ increases
\begin{equation}
	F( \rho_{N} , \rho_{\infty}) \rightarrow 1.
\end{equation}
\end{theorem}
This tells us that, assuming a suitable reference exists, the convergence behaviour of the operators $\tau_{N}$ is inherited by the physical states, $\rho_{N}$.  In Ref.~\cite{Campbell12} a similar result for the case $\tau_{\mathrm{ref}}=\tau_{\infty}$ was shown. Although this is useful in some cases, often $\tau_{\infty}$ will not always satisfy the conditions for a reference state and so this result allows us to use another operator as a proxy. 

Our approach to the proof is to find $\tr ( \Pi \rho_{N} )$ by calculating the expectation value of $\tau_{N}$ with respect to $\Pi^{-1}$. These quantities are related by
\begin{equation}
	\tr( \Pi^{-1} \tau_{N} ) = \frac{\tr (  \Pi \Pi^{-1} \rho_{N}) }{\tr ( \Pi \rho_{N})} = \frac{1}{\tr( \Pi \rho_{N})} .
\end{equation}
The Gaussian filter can always be written as the exponential of some Hamiltonian 
\begin{equation}
	H_{\Pi}=\sum_{j} \beta_{j} V \hat a_{j}^{\dagger} \hat{a}_{j}V^{\dagger} ,
\end{equation}	
such that $\Pi = \exp(- H_{\Pi})$, where $H_{\Pi}$ is Hermitian and quadratic in annihilation/creation operators.  The inverse filter is then $\Pi^{-1}=\exp(+H_{\Pi})$ and
\begin{equation}
	\tr( \Pi^{-1} \tau_{N} ) = \tr \left( \sum_{k=0}^{\infty} \frac{H_{\Pi}^{k}}{k!}  \tau_{N} \right) .
\end{equation}
Each term is a sum of moments of degree $2k$ so it is tempting to think that Thm.~(\ref{momentsConverge}) can be directly applied.  However, the whole sum has infinitely many terms so Thm.~(\ref{momentsConverge}) is not applicable.  Each $\tr ( H_{\Pi}^{k}  \tau_{\mathrm{ref}} ) $ is positive and, by assumption, the infinite sum gives a finite value.  It follows that for any $\epsilon>0$ we can pick an integer $k_{c}$ such that the truncation satisfies
\begin{equation}
\label{tail}
	| \tr \left( \sum_{k=k_{c}+1}^{\infty} \frac{H_{\Pi}^{k}}{k!}  \tau_{\mathrm{ref}} \right) | < \epsilon ,
\end{equation}
for the reference state $\tau_{\mathrm{ref}}$.  Furthermore, using this $k_{c}$, we can partition the summation for $\tau_{N}$ such that
\begin{equation}
	\tr ( \Pi^{-1} \tau_{N} ) = \tr \left( \sum_{k=0}^{k_{c}} \frac{H_{\Pi}^{k}}{k!}  \tau_{N} \right) +  \tr \left( \sum_{k=k_{c}+1}^{\infty} \frac{H_{\Pi}^{k}}{k!}  \tau_{N} \right) .
\end{equation}
Now, the first term is a finite sum and so the results of Thm.~(\ref{momentsConverge}) do apply to this portion of the sum.  Hence, for sufficiently large $N$
\begin{eqnarray}
	| \tr ( \Pi^{-1} (\tau_{N}-\rho_{\infty}) ) | & \leq &  \epsilon + |  \tr ( \sum_{k=k_{c}+1}^{\infty} \frac{H_{\Pi}^{k}}{k!}( \tau_{N} - \tau_{\infty}) ) |, 
	\nonumber	
	\\ 	 
	& \leq &  \epsilon + 2 |  \tr ( \sum_{k=k_{c}+1}^{\infty} \frac{H_{\Pi}^{k}}{k!} \tau_{\mathrm{ref}} ) | ,
\end{eqnarray}
where in the last line we have used the properties of a reference state.  Combining this with (\ref{tail}) we deduce that for large enough $N$
\begin{equation}
	| \tr ( \Pi^{-1} (\tau_{N} - \rho_{\infty}) ) |  \leq  3 \epsilon .
\end{equation}
By taking longer truncations $k_{c}$ and larger $N$, the value of $\epsilon$ can be made arbitrarily small.  Therefore, we have that for increasing $N$
\begin{equation}
	 \tr ( \Pi^{-1} \tau_{N}  ) \rightarrow \tr ( \Pi^{-1} \rho_{\infty})  .
\end{equation}
Consequently, $\tr ( \Pi \rho_{N}  ) $ approaches $\tr ( \Pi \rho_{\infty}  )$ and the fidelity between these states approaches unity.

These techniques, in particular the use of reference states, give us a handle on this difficult part of the analysis.   The central limit theorems ensure that the filtered operators converge to a Gaussian.  However, alone, the central limit theorems provide no guarantees on the behavior of expectation values for unnormalizable operations like $\Pi^{-1}$.  Indeed, it is easy to find central limit sequences for which $\tr ( \Pi^{-1} \tau_{N} )$ diverges with $N$.  In such pathological examples, the physical states $\rho_{N}$ would also diverge with ever increasing energy.  However, in light the arguments presented, when a suitable reference state exists these pathologies cannot occur.

The limiting operator $\tau_{\infty}$ may sometimes be chosen as a reference state, but in some cases it is unsuitable.  Now we will discuss a few facts that simplify the task of finding a suitable reference state.  First,  we note that if
\begin{equation}
		| \tr ( H^{(k)} \tau_{\mathrm{ref}} ) |  =  \tr ( H^{(k)} \tau_{\mathrm{ref}} )  
\end{equation}
holds for all second moments then it must hold for all higher moments also.  By Wick's theorem (see App  \ref{APPmoments}) the higher moments for Gaussian states are simply a positive polynomial in $2^{\mathrm{nd}}$ moments.  Consequently, positivity of higher moments is inherited from positivity of second moments, which simplifies the search for appropriate reference states.   

For single mode states there is one very simple class of potential reference states.  Consider the pure squeezed states
\begin{equation}
	\ket{\psi_{R}} = \sum_{n=0}^{\infty} \lambda^{n} \ket{2n},
\end{equation}
where $0<\lambda<1$ and $|\lambda|=\lambda$.  Calculating $\bra{\psi_{R}} \hat a^{\dagger}\hat{a} \ket{\psi_{R}}$ and 
$\bra{\psi_{R}}\hat a \hat a \ket{\psi_{R}}$ we find they are real, positive, and increasing with $\lambda$.  These form a promising class of single mode reference states, as for any $\tau_{1}$ and any even moment $H^{(k)}$ we can find a large enough $\lambda$ such that $|\tr( H^{(k)}\tau_{1})| < \tr( H^{(k)}\tau_{\mathrm{ref}}) $.  However, Thm.\ \ref{Strong} also requires that $\tr ( \Pi^{-1} \tau_{\mathrm{ref}} ) < \infty $, but there will be a critical value of $\lambda$ at which this expectation value diverges.  For many single mode central limit sequences there will exist a choice of $\lambda$ that satisfies both these requirements, though some counterexamples do exist.  For multi-mode problems, pure squeezed or entangled state can make suitable reference states.

Above, we focused on the even moments of the Gaussian state.  It is easy to check that all Gaussian states, with zero first moments, have vanishing odd moments.  This seems to entail severe constraints on the odd moments of $\tau_{1}$.  However, this problem can be remedied by a physical procedure that is a CV version of twirling.   The concept of twirling, arising also in entanglement distillation of finite dimensional systems \cite{BBPSSW01a} and magic state distillation \cite{BraKit05}, generates a symmetry in the initial resource.  This symmetry significantly simplifies the analysis of a protocol's convergence.  The twirling map we prescribe here applies, with 50/50 probability, either the identity or the local Gaussian unitary, $U_{T}$, that maps $\hat{a}_{j} \mapsto - \hat{a}_{j}$ for all $j$, such that
\begin{equation}
	\mathcal{T}(\rho_{1}) = \frac{1}{2}( \rho_{1} + U_{T} \rho_{1} U_{T}^{\dagger}) .
\end{equation}
For such a twirled state, the odd moments have zero expectation value whereas the even moments are unchanged.  Furthermore, twirling the physical state also results in twirling on the level of the filtered object, since
\begin{eqnarray}
	\frac{P \mathcal{T}(\rho_{1}) P}{\tr [  P \mathcal{T}(\rho_{1}) P ] } & = & \frac{ \mathcal{T}(P \rho_{1}P ) }{\tr [  \mathcal{T}(P \rho_{1} P) ]}   \\  \nonumber
	& = &  \frac{ \mathcal{T}(P \rho_{1}P ) }{\tr [ P \rho_{1} P ]}  = \mathcal{T}( \tau_{1} ) .
\end{eqnarray}
The above follows immediately from the observation that $U_{T}$ commutes with $P$ as it does not change second moments.  Another consequence of twirling preserving second moments is that the central limit sequence evolves to the same $\tau_{\infty}$ independently of whether we twirled or not.  However, having twirled and eliminated all odd moments makes it possible for good reference states to exist and for Thm.\ \ref{Strong} to hold.

Finally, we give another remark on condition (i) of the definition of reference states.  For brevity we stated that this must hold for all products of operators $\{ \hat{b}_{j}^{\dagger}, \hat{b}_{j} \}$.  However, we only need to verify that the condition is valid for all \textit{normally} ordered operators. Recall that normally ordered operators have all $\hat{b}_{j}^{\dagger}$ operators to the left side of any $\hat{b}_{j}$ operators, so $\hat{b}^{\dagger}_{j}\hat{b}$ is normally order but $\hat{b}_{j} \hat{b}^{\dagger}_{j}$ is not.    By using $\hat{b}_{j}\hat{b}_{j}^{\dagger}= \hat{b}_{j}^{\dagger}\hat{b}_{j} + \id$ it is easy to rewrite the relevant operators --- those composed of products from the set $\{ \hat{b}_{j}^{\dagger}, \hat{b}_{j} \}$ --- as a positive sum of normally ordered operators.  Provided $|\tr(H^{(k)} \tau_{1})| \leq \tr(H^{(k)}\tau_{\mathrm{ref}})$ for normally ordered operators, it follows that the same holds for positive sums of normally ordered operators.  Again, this observation is useful for reducing the workload of verifying that a purported reference state indeed meets all the requirements.

\section{Hybrid (continuous variable) quantum repeaters}

Quantum repeaters are one of the main applications of the various variants of entanglement distillation discussed and proposed here. The aim of quantum repeaters is to distribute entanglement, despite the presence of noise, over large distances.  There are many variants of such schemes, though they all share the common feature of using entanglement swapping rounds that entangle pairs that have not interacted in the past and distillation to reduce noise.  It is long established \cite{DBCZ01a,DLCZ01a} that discrete variable repeater networks can achieve distances far beyond those feasible by direct transmission of quantum states.  Despite considerable work on CV entanglement distillation, it has, surprisingly, not been shown that CV repeater networks can outperform direct transmission.  Here, we give the first evidence that CV repeater networks can outperform direct transmission, albeit under some idealised conditions.  In particular, we do not compute rates of entanglement production as these calculations are very computationally intensive for CV systems and so beyond our scope.

\subsection{Primitives}

The primitives discussed and introduced here are useful in constructing CV quantum repeater schemes.  It is beyond the scope of the present work to present a comprehensive study of the possible repeater schemes that can be devised based on these basic elements. Given the importance of this application, we however sketch what parameters may be varied in variants of such schemes.
\begin{itemize}
\item {\it Gaussification:} There are several conceivable ways of performing Gaussification, 
including a recursive-Gaussifier, a temporally-reordered recursive-GaussiÞer, a 
pumping-Gaussifier, and others.  Since convergence of these protocols is fast, and in order not to
arrive at low rates, it seems advisable to perform very few steps in each instance. The resource requirements, 
in particular involving memory requirements, are different in these schemes. The framework developed here and in 
Ref.\ \cite{Campbell12} allows for a trade-off between success probability and quality of the output, when projecting onto
Gaussian states different from the vacuum.
\item  {\it Swapping:} The precise procedure of entanglement swapping may be varied, with the original nested scheme being only
one possibility. For Gaussian states, the optimum Gaussian entanglement swapping scheme is known \cite{Ohliger,Obermaier}
and is used subsequently. But other swapping steps are conceivable as well, such as mixing inputs at a symmetric beam splitter and projecting the outputs onto certain photon number states.
\item  {\it Non-Gaussian operations:} 
Given a source of Gaussian entangled states some non-Gaussian operation will be required prior to Gaussification, which is said to de-Gaussify the initial state. There are many possible ways to perform non-Gaussian operations in the scheme, such as in particular only at the beginning, or also in later steps of the protocol. Also, several kinds of non-Gaussian steps have been considered in the literature so far.
This includes (i) a mixing of the signal at a beam splitter with a single photon state, followed by a measurement at 
one of the output ports \cite{Browne03,Eisert04}. We will refer to this step as single photon replacement since a single photon is both added and removed. (ii) One can think of photon subtraction schemes,  again leading to non-Gaussian states \cite{Subtraction1,Subtraction2,Subtraction3,Fiur10,Eisert04,Kim08}. 
(iii) Ref.\ \cite{Fiur10} introduces a modified non-Gaussian operation that is 
experimentally more challenging, but suggests better purification. 
\item  {\it Non-Gaussian inputs:} In order to arrive at reasonable success probabilities, it may also be advantageous to make use of
non-Gaussian input states that have higher photon numbers suppressed by their very preparation mechanism.  For example, using entangled pairs generated from quantum dots in bi-photon cascades (see, e.g., Ref.\ \cite{BiPhotons}).
\end{itemize}
These parameters can be altered in benchmarking the functioning of such protocols, along the lines as has recently
been done for discrete-variable quantum repeater schemes \cite{BrussRepeater}. Needless to
say, in any such effort, not only the losses in transmission have to be taken into account, but also the 
impact of imperfect swappings and Gaussification as well as issues of mode matching.
Symmetric entanglement distillation schemes may be also favourable compared to 
asymmetric schemes \cite{Lund}. 

% Note that Gaussification as such --- deriving from an argument presented in Ref.\cite{Fiurasek} --- has the potential to distribute entanglement over arbitrary distances. 

\subsection{Our repeater network}

\begin{figure}
\includegraphics{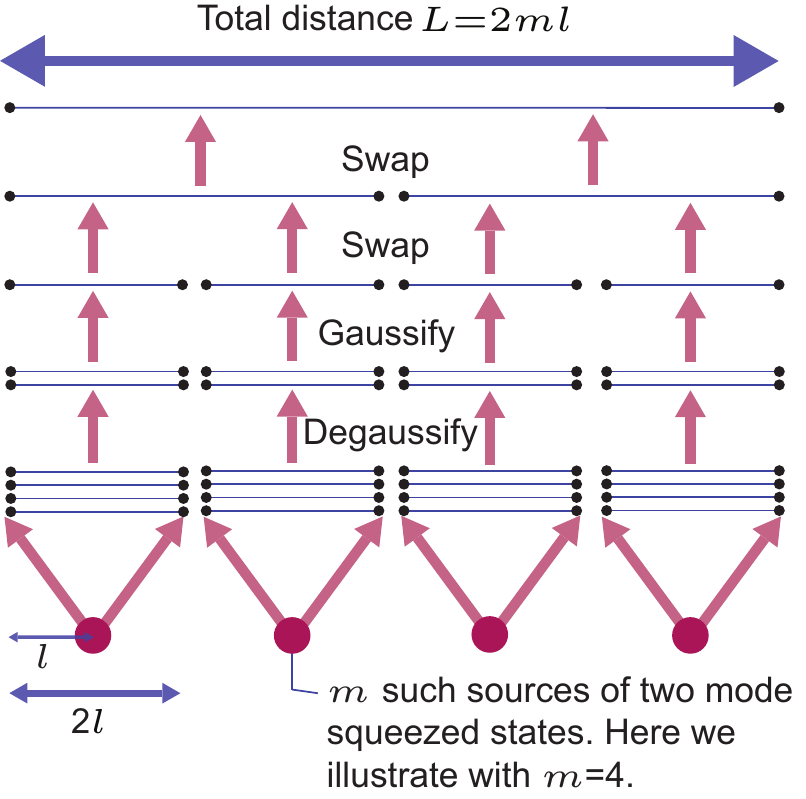}
\caption{A schematic of the CV repeater network considered here.  Sources are assumed to produce pure two-mode squeezed state of some chosen squeezing.  Channels are predominately affected by attenuation, but also a small amount of room temperature thermal noise.  Degaussification is performed by photon replacement.  Gaussification is performed as described here and in previous work using measurements projecting onto the vacuum, 
using many copies of the de-Gaussified state and asymptotically approaching a Gaussian state.  Entanglement swapping uses deterministic optimal continuous swapping protocol.}
\label{RepeaterNetwork}
\end{figure}

Here, we introduce a concrete class of quantum repeaters than are analyzed in the next section.  In these protocols, any covariance matrix of any two-mode Gaussian state $\rho$ encountered at any step is of the form
\begin{equation}\label{CM2}
	\Gamma_\rho = \left( \begin{array}{cccc}
	C & 0 &S & 0 \\
	0 &C & 0 &-S \\
	S & 0 &C & 0 \\
	0 &-S & 0 &C \\
	\end{array} \right)  ,
\end{equation}
where $C,S\geq 0$ with $C^2\geq 1+S^2$. For a pure two-mode squeezed state, $C^2= 1+S^2$, this equality
becoming an inequality in case of mixed Gaussian states. The EPR uncertainty \cite{Duan}, which for a Gaussian state with a covariance matrix
as in Eq.\ (\ref{CM2}) takes the simple form
\begin{equation}
	\Delta(\rho) = C-S.
\end{equation}
Rates in CV key distribution schemes will, in particular, relate to the above quantity. Indeed, a Gaussian 
state $\rho$ with a covariance matrix of the above form is entangled if and only if $\Delta(\rho) <1$
(the implication still being valid in one direction for non-Gaussian states).

The numerics presented here are based on the following CV repeater protocol (also illustrated in Fig.~\ref{RepeaterNetwork}):
\begin{enumerate}
	 \item Each of the $m=2^k$ sources repeatedly produce many copies of a pure two mode squeezed state (squeezing parameter $r$); 
	\item  \earl{Each half of every entangled pair is transmitted a distance $l$ to a repeater node, and so becomes noisy due to attenuation;}
	\item  Photon replacement is used to probabilistically de-Gaussify; 
       \item  The de-Gaussified states are now iteratively Gaussified; 
       \item  The Gaussified state are swapped $k$ times until a entangled is shared across the full distance $L=2ml = 2^{k+1}l$.       	  
\end{enumerate}
We require that the first step produces pure two-mode squeezed states of the form of Eq.~(\ref{CM2}).  We set $C=\cosh (2r)$ and $S=\sinh (2r)$ and call $r>0$ the squeezing parameter, for which we consider a range of possible values.

\earl{After the second step, the entangled pairs suffers noise from transmission over a lossy channel, becoming mixed states prior to distillation.}  For photons traveling in optical fiber the dominant noise source is attenuation through absorption, scattering, and mode mismatching.  Indeed, attenuation is so dominant that previous analysis of CV distillation protocols has focused on pure attenuation noise channels.   A solely attenuating channel will never completed eliminate the entanglement of a transmitted two-mode squeezed state.  We consider Gaussian channels with a small contribution of additional noise, on top of attenuation, such that covariance matrices evolve as
\begin{equation}
	\gamma \mapsto  e^{- l / l_{{\rm att}}} \gamma +(1+ 2 n_{\mathrm{th}})(1-e^{- l / l_{{\rm att}}} ) \id ,
\end{equation}
where $l$ is is the distance (herein all distance in $km$) travelled by each mode and $l_{\mathrm{{\rm att}}}$ is the attenuation length of the fiber optic.  In the infinite distance limit the state becomes thermal with an average photon number $n_{\mathrm{th}}$.  Applying such a noise model to the pure Gaussian state of Eq.~(\ref{CM2}) gives a mixed state of a similar form where
\begin{eqnarray}
      C  & =  &  e^{- l / l_{{\rm att}}} \cosh ( 2r )  +  (1+ 2 n_{\mathrm{th}})(1-e^{- l / l_{{\rm att}}} ) , \\ \nonumber
      S & = &  e^{- l / l_{{\rm att}}} \sinh ( 2r ) .
\end{eqnarray}
Herein we take $l_{\mathrm{{\rm att}}}=22$ km as this is the state of the art for current fiber optic cable.  For a pure attenuation channel $n_{\mathrm{th}}=0$,  but we take $n_{\mathrm{th}}= 10^{-8}$ as this corresponds to the thermal photon occupation at room temperature.  The interesting feature of our analysis is that this modest additional noise source is sufficient to put a hard cap on the distance at which various protocols can propagate entanglement.  Assuming an initially pure two-mode squeezed state with squeezing parameter $r$, the maximum distance possible by direct transmission before the state is separable is easily found to be
\begin{equation*}
l_{\mathrm{max}}(r) = 2  l_{\mathrm{{\rm att}}} \ln \left( \frac{1 + 2 n_{\mathrm{th}}  - \cosh (2r) + \sinh (2r)}{2 n_{\mathrm{th}}}  \right) .
\end{equation*}
This increases with $r$ approaching the limiting value
\begin{equation}
 \lim_{r \rightarrow \infty} l_{\mathrm{max}}(r) = 2  l_{\mathrm{{\rm att}}} \ln \left( 1 + \frac{1}{2 n_{\mathrm{th}}}   \right)  ,
\end{equation}
which for our chosen parameters evaluates to $780$ km.  \earl{Recent continuous-variable experiments have achieved quantum cryptography,  directly and without aid of repeaters,  at a distance of $80$ km~\cite{Repeater80km}}. Our upper bound is of \earl{roughly} the same order of magnitude, but larger as we take an optimistic noise model.  \earl{We present results on two variants on the noise model.  Analysis (i), as presented in Fig.~(\ref{Fig_RepeaterDist}i), assumes that transmission noise dominates all other noise sources.  Analysis (ii), as presented in Fig.~(\ref{Fig_RepeaterDist}ii), is more pessimistic and assumes that an additional $50\%$ photon loss occurs within the repeater station.   This additional loss equates to over 15km of optical fibre, but can also be attributed to other effects such as mode mismatching and detector inefficiencies.}   

% We do not take into account additional losses due to imperfect swapping or Gaussification; this analysis is done to merely exemplify the optimum performance anticipated, to be used in benchmarking efforts \cite{Benchmarking} of more realistic schemes. 

On the 3rd step of our repeater protocol we de-Gaussify by using symmetric photon replacement.  The process begins with mixing a mode of the entangled pair on beam-splitter of transmittivity $\eta^{2}\in[0,1]$ where the second input mode contains a single photon.  Next, the reflected signal mode is measured with a single photon resolving detector and we postselect on seeing a single photon.  Such de-Gaussification procedures have been extensively studied~\cite{Browne03,Eisert04,Lund} so we shall not repeat a full analysis here.  However, it is informative to introduce the variable
\begin{equation}
	\epsilon ( \rho ) = \frac{ \bra{1,0} \rho \ket{ 1,0} }{   \bra{ 1,1} \rho \ket{0,0} } ,
\end{equation}
which is meaningful because it is unchanged by photon replacement, or indeed any operation with Kraus operators diagonal in the Fock basis.  In particular, for a symmetric Gaussian state of form Eq.~(\ref{CM2}) we find
\begin{equation}
	\epsilon ( \rho ) = \frac{C^{2}-S^{2}-1}{2 S} .
\end{equation}
This variable is of interest as it cannot be increased either by Gaussification \footnote{Here we assume Gaussification using $P=\kb{0,0}{0,0}$, although the increase of $\epsilon$ by Gaussification using different choices of $P$.} or photon replacement.  Indeed, $\epsilon$ remains unchanged by any local de-Gaussifying procedure resulting in Kraus operators diagonal in the Fock basis.

In a variation of the argument presented in Ref.\ \cite{Lund} to accommodate for thermal noise, one obtains that the net effect of de-Gaussification and subsequent Gaussification, using $P=\kb{0,0}{0,0}$, is that the state evolves to a Gaussian with
\begin{eqnarray}
	C & = & \frac{   \Lambda^2(1 - \epsilon^2) + 1 }{(1 - \epsilon \Lambda)^2 - \Lambda^2}  , \\ \nonumber
	S & = &  \frac{2 \Lambda}{(1 - \epsilon \Lambda )^2 - \Lambda^2} ,
\end{eqnarray}
where $\epsilon$ depends on $\rho$ after transmission through the noise channel and $\Lambda$ can be tuned to any value in the interval $0 < \Lambda < (1 + \epsilon)^{-1}$ by suitable choice of the beam-splitter transmittivity used in de-Gaussification.    Larger values of $\Lambda$ provide more entanglement in the final state,  and we have numerically found that larger values also produce repeater networks capable of reaching larger distances.  However, larger values of $\Lambda$ also significantly 
reduce the success probability of de-Gaussification.  Herein we assume that $\Lambda = 0.99/(1+\epsilon)$, as any further increase results in only a negligible increase in maximum repeater distance.

Having distributed entanglement and distilled at repeater stations, in the last step we perform swapping operations to generate entanglement between the most distant repeater nodes.  In order to describe the optimum Gaussian entanglement swapping \cite{Obermaier,Ohliger} consider the
function $g: \R^2\rightarrow \R^2$, defined as
\begin{equation}
	g(x,y)= \left( x-\frac{y^2}{2x}, \frac{y^2}{2x} \right).
\end{equation}
Indeed, the covariance matrix before of the form (\ref{CM2}) with $C,S\geq 0$ is mapped onto one of the same form with
\begin{equation}
	(C',S')= g(C,S).
\end{equation}
% (As mentioned before, even for Gaussian states, non-Gaussian swaps may be beneficial to the scheme.)
If $2l$ is the distance between the repeater stations, such a scheme would distribute an entangled state over a physical distance of 
$l 2^{(k+1)}$ for $k$ swaps.  As such repeater networks are typically divided into $m=2^{k}$ intervals for some integer $k$.   This results in a mapping $g^{k}$.

\subsection{Maximum distance of repeater networks}

\begin{figure*}
\includegraphics{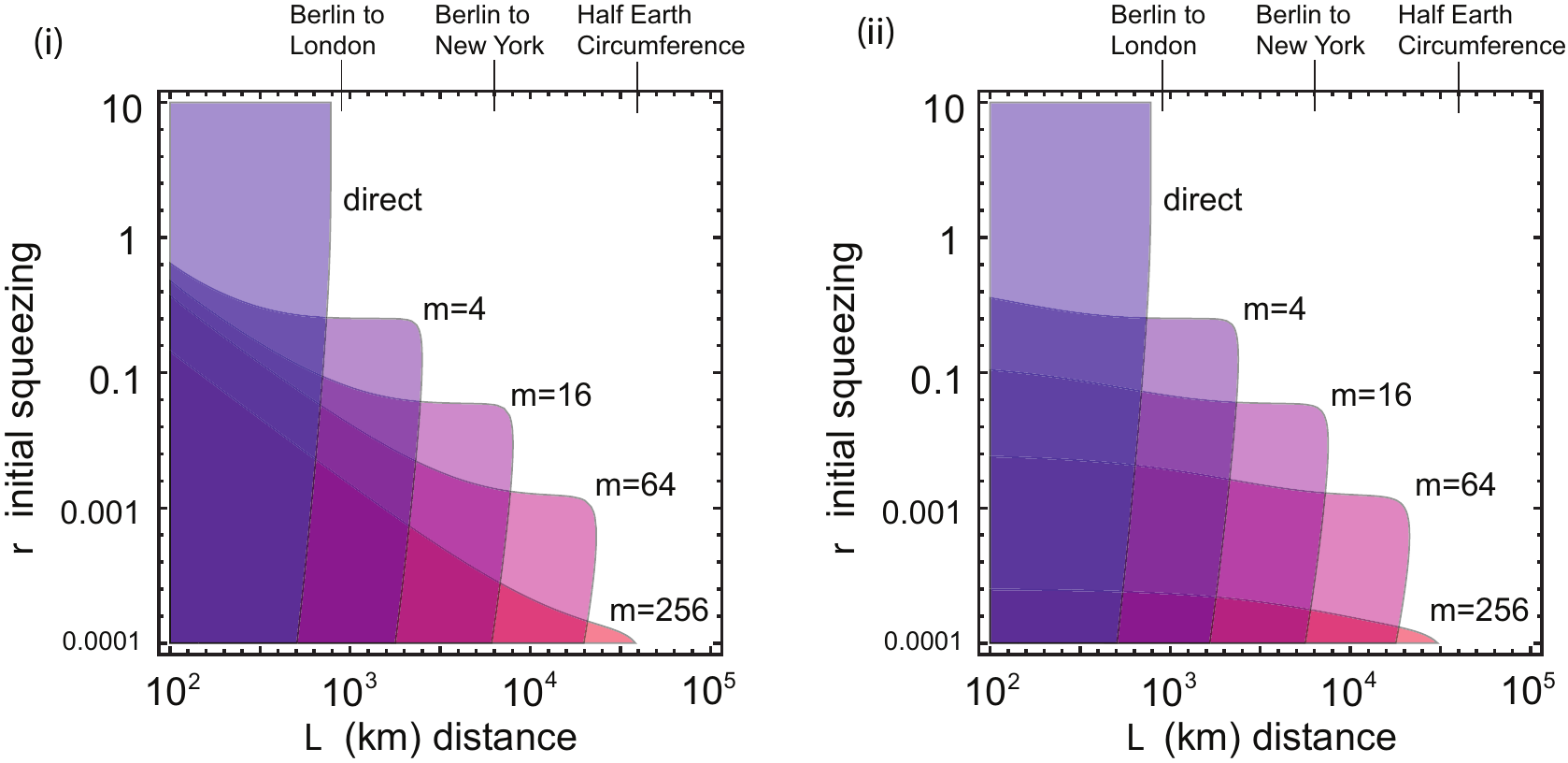}
\caption{The maximum attainable distance, $L$, for a range of initial squeezing, $r$, for which entanglement can be distributed.  \earl{In plot (i) noise arises wholly from transmission between repeater stations, whereas in plot (ii) we incur an addition 50\% attenuation within each repeater station (see text for more details)}.  Regions are shown for direct transmission and CV quantum repeater networks dividing the distance up into $m$ intervals.  Some noteworthy terrestrial scales are shown at the top.}
\label{Fig_RepeaterDist}
\end{figure*}

We now discuss the maximum distance that can be reached in the repeater scheme outlined in the previous section.  We say a scheme achieves a distance, $L$, whenever it produces an entangled state, as verified by the Duan criteria, $\Delta(\rho)<1$, between the distant repeater nodes.   These results are summarised by Fig.~\ref{Fig_RepeaterDist}, where we show the achievable distances for different numbers of repeater stations and a range of initial squeezing parameters.  For direct transmission --- where no actual repeater techniques are exploited --- we find performance is best in the large squeezing regime.  However, we see that by using more repeater stations, and hence more intensive distillation, greater distances may be achieved.  This provides the first evidence that CV techniques may achieve distances of a global scale, whereas direct transmission is incapable of achieving relatively short distances, such as Berlin to London.   \earl{Comparing analysis (i) and (ii), the additional noise of the latter model does slightly reduce the maximum distance, but the decrease is very small.  We also see that using more repeater stations typically requires a smaller initial squeezing, and this effect is more pronounced in analysis (ii) for short distances.}  This is consistent with observations made in Ref. \cite{Lund} where they observed that distillation was more effective when combined with smaller initial squeezing.  A possible explanation for this feature is that the more squeezed the initial state the more mixed the final state after suffering photon loss. Furthermore, the Gaussification process, while increasing entanglement, does not actually increase the purity, so limiting the impact of photon loss on purity is the key parameter to be optimised.

These results can be contrasted with those of Ref. \cite{Obermaier}. Its authors compared the performance of direct transmission to use of entanglement swapping, though without the benefits of any entanglement distillation, and found direct transmission to be preferable.   Our noise model and figure of merit differs from those of Ref.~\cite{Obermaier}, but our own numerics also found that entanglement swapping without distillation always achieves significantly inferior distances.   Such behaviour is a unique feature of CV protocols as the discrete variable protocol of Ref.~\cite{DLCZ01a} shows that swapping, albeit with some postselection, can be beneficial.  We also considered some other variants of our repeater network.  For instance, we considered  several \textit{nested} repeater schemes, where each entanglement swap is interleaved with distillation.  Again, we found that these alternative protocols achieved shorter maximum distances compared to the protocol explicitly described in the previous section. 

It seems that the results in Fig.~\ref{RepeaterNetwork}, at least using the specific forms of Gaussification and de-Gaussification considered here, show the upper bounds of what is feasible with current technology.  However, this does leave open the possibility of using alternative de-Gaussification procedures, such as that proposed in Ref.~\cite{Fiur10}, or suitable deterministically prepared non-Gaussian states to start with.  
As commented earlier, the parameter $\epsilon$ is non-increasing through our distillation techniques, though those techniques 
can vary this parameter potentially leading to an increase of the maximum attainable distance. However, to-date such proposals are even more technologically challenging than replacement of a single photon.  On the other hand, while CV systems pay a high price for de-Gaussification they can produce two mode squeezed states at intrinsically higher rates than single photon sources. They also benefit from the higher efficiency of homodyne detectors.  In future work, a careful analysis of rates will be made, including also a comparison with common discrete variable schemes that weighs these relative merits.

\section{Summary and conclusion}

In this work, we have further introduced and elaborated upon a formalism general enough to capture all of  the known schemes of entanglement distillation leading to Gaussian quantum states, as well as to construct a plethora of new ones. The flexibility of the approach allows trading success probabilities against the quality of the resulting entangled states, or to realistically
take memory requirements into account. As such, the formalism presented here provides a natural starting point for 
comprehensive comparisons of different entanglement distillation schemes in the continuous-variable setting. At the roots
of the formalism is a novel kind of non-commutative central limit theorem that is laid out in great detail. We also discuss the 
implications of the findings for devising novel schemes for quantum repeaters and highlight both potential and limitations.
It is the hope that the general framework developed here gives a basis for assessing to what extent experimental
large distance continuous variable quantum communication is truly feasible.

\section{Acknowledgements}
This work has been supported by the BMBF (QuOReP), the EU (Q-ESSENCE), the ERC (TAQ), the EURYI 
and the EPSRC.  We thank A. Serafini and M. Paris for interesting discussions that contributed ideas toward the design of the pumping-Gaussifier.

\appendix

\section{Conjugation lemma}
\label{App_Comm_Lemma}

Here, we show that for any Gaussian operator, $P$, with zero first moments, we have $U (P \ox P) U^{\dagger}= (P \ox P) $ where $U$ is a multi-lateral beam splitter transformation.  The Gaussian operator can be expressed as $P \ox P = k \exp ( -  ( H_{A} + H_{B}) )$ and 
\begin{equation}
	H_{X=A,B}= \sum_{i,j} h_{i,j} \hat{Q}_{Xi} \hat{Q}_{Xj},
\end{equation}
for some $h_{i,j}$ and so
\begin{equation}
	H_{A}+H_{B}= \sum_{i,j} h_{i,j} ( \hat{Q}_{Ai} \hat{Q}_{Aj}+ \hat{Q}_{Bi} \hat{Q}_{Bj} ) .
\end{equation}
The beam-splitters cause
\begin{equation}
	U  (P \ox P) U^{\dagger} = k \exp ( -  U( H_{A} + H_{B} )U^{\dagger}),
\end{equation}
and so we simply need to show that
\begin{equation}
	U(  \hat{Q}_{Ai} \hat{Q}_{Aj}+ \hat{Q}_{Bi} \hat{Q}_{Bj}  )U^{\dagger} =  \hat{Q}_{Ai} \hat{Q}_{Aj}+ \hat{Q}_{Bi} \hat{Q}_{Bj}  .
\end{equation}
Using the shorthand $J_{i,j}$ for the R.H.S and conjugating the quadrature operators with the unitary we have
\begin{eqnarray*}
	UJ_{i,j}U^{\dagger} & = &  ( \sqrt{T}  \hat{Q}_{Bi}+ \sqrt{R}  \hat{Q}_{Ai}  ) ( \sqrt{T}  \hat{Q}_{Bj}+ \sqrt{R}  \hat{Q}_{Aj}  ) \nonumber\\
	& & +   ( \sqrt{T}  \hat{Q}_{Bi} - \sqrt{R}  \hat{Q}_{Ai}  ) ( \sqrt{T}  \hat{Q}_{Bj}  - \sqrt{R}  \hat{Q}_{Aj}  )   .
\end{eqnarray*}
Expanding out, we find the cross terms ($ \hat{Q}_{Ai} \hat{Q}_{Bj}$ and $ \hat{Q}_{Bi} \hat{Q}_{Aj}$) cancel leaving only
\begin{eqnarray}
	UJ_{i,j}U^{\dagger} & = & (R+T)(  \hat{Q}_{Ai} \hat{Q}_{Aj}+ \hat{Q}_{Bi} \hat{Q}_{Bj}) .
\end{eqnarray}
Recalling $R+T=1$, we have $UJ_{i,j}U^{\dagger}=J_{i,j}$, which in turn entails the result  $U (P \ox P) U^{\dagger}= (P \ox P) $.

\section{An inequality}
\label{miniIdentity}

For any complex $a$ and $b$ satisfying $|a| \leq 1$ and $|b|\leq 1$ and any integer $N$, we have $|a^{N}-b^{N}|\leq N |a-b|$.  For $N=1$ it is trivial and for higher $N$ it is proven iteratively,
\begin{eqnarray} 
	| a^{N}-b^{N} | & = &  |(a-b) a^{N-1} + b(a^{N-1}-b^{N-1}) | \\ \nonumber
	& \leq & |a-b| + | a^{N-1}-b^{N-1}|,
\end{eqnarray}
where we have used the triangle inequality and $|a^{N-1}|\leq 1$ and $|b| \leq 1$.  Each unit increase in $N$ contributes at most an additional $|a-b|$, and so we have the desired result.

\section{Uniform convergence}
\label{APPuniform}

In the main text we prove Thm.\ \ref{CLTchar} for an individual point of phase space.  Here, we extend it to a uniform result over balls of finite radius.  For any finite set of points $\mathcal{R}_{\mathrm{finite}}=\{\vec{r}_{1}, \vec{r}_{2},\dots \}$ convergence is uniform over that set as it is bounded by the point that converges slowest.  For any small distance, $\delta$, we can find a $\mathcal{R}_{\mathrm{finite}}$ such that any point inside the ball is less than distance $\delta$ from some point in the finite set.  All $\chi \in \{ \chi_{\tau_{N}} \}$ are continuous and within the ball there is a maximum possible gradient.  Hence, for every point in the ball we can approximate the characteristic function by a nearby point in the finite set $\mathcal{R}_{\mathrm{finite}}$ and uniform convergence follows.  

\section{Moments convergence}
\label{APPmoments}

Here, we present a proof of Thm.\ \ref{momentsConverge} that follows the combinatorial argument of Refs.\ \cite{Giri78,Petz90}.  Our proof is not as general, but benefits from requiring less mathematical background.  We consider a single $k^{\mathrm{th}}$ moment $H^{(k)}=\prod_{j} H_{j}$,
\begin{eqnarray}
	\tr ( H^{(k)} \tau_{N} ) & = & 	 \tr \left[  \left( \prod_{j} H_{j} \right) U \tau_{1}^{\otimes  N} U^{\dagger} \right] , \\ \nonumber
	 & = &\frac{1}{N^{k/2}} \tr \left[ \prod_{j} \left(  \sum_{x=1,\dots,N}  H_{j,x}   \right)  \tau_{1}^{\otimes  N} \right] ,
\end{eqnarray}
where $H_{j,x}$ indicates the $H_{j}$ operator but acting on the $x^{\mathrm{th}}$ of the $N$ systems.   We need to expand out the brackets and some way of labelling terms.   We have $k$ different operators that can act on $N$ different copies.  Each possibility can be represented by a partition of $k$ values in to $N$ bins.  For example, for $k=4$ and $N=5$ a possible partition is $B=\{ \{ 1,2 \} ,\{ \}, \{ 3 \}, \{ 4\}  , \{ \}\}$ with which we associate with a term $ H_{1}H_{2} \ox \id \ox H_{3} \ox H_{4} \ox \id$.  In general, for a partition $B=\{ B_{1}, B_{2}, B_{3},\dots, B_{N}\}$ we associate an operator
\begin{equation}
	 H_{B} = \ox_{x=1}^{N}   H_{B_{x}} ,
\end{equation}
where
\begin{equation}
	H_{B_{x}} = \prod_{j \in B_{x}} H_{j} ,
\end{equation}
with the product over $j \in B_{x}$ always taken in order of smallest to largest value of $j$.  In this notation
\begin{eqnarray}
	\tr ( H_{B} \tau_{N} ) & = &\frac{1}{N^{k/2}} \tr \left[ \left( \ox_{x=1}^{N} H_{B_{x}}  \right) \tau_{1}^{\otimes  N} \right] , \\ \nonumber
	 & = &\frac{1}{N^{k/2}} \prod_{x=1}^{N} \tr ( H_{B_{x}}  \tau_{1} ) .
\end{eqnarray}
The next key step of the proof is a smart way of collecting up terms with similar properties.  We define $L(B)$ to be the number of non-empty bins in $B$ and then collect terms with the same value.
\begin{eqnarray}
	\tr ( H^{(k)}  \tau_{N} ) & = &  	 \frac{1}{N^{k/2}} \sum_{b }  \sum_{L(B)=b} \tr \left[ H_{B}  \tau_{1}^{\otimes N} \right]\nonumber \\ 
	& = &	 \frac{1}{N^{k/2}} \sum_{b }  \sum_{L(B)=b} \prod_{x=1}^{N} \tr ( H_{B_{x}}  \tau_{1} ) .
\end{eqnarray}
For any $B$ there are $N! / (N- L(B) )! = N(N-1) \cdots (N-L(B)+1)$ partitions that differ by only a permutation of whole bins.  For instance, $B=\{ \{ 1,2 \} ,\{ \}, \{ 3 \}, \{ 4\}  , \{ \}\}$ and $B'=\{ \{ \}, \{ 3 \}, \{ 1,2 \} , \{ 4\}  , \{ \}\}$ differ only by a permutation of whole bins, and so give the same expectation value. We can also choose a canonical set $\mathcal{B}$ such that for every $B$ there exists a unique $B' \in \mathcal{B}$ such that $B$ and $B'$ differ only by a permutation of whole bins.  By summing over just the canonical set we have,
\begin{equation}
	\tr ( H^{(k)}  \tau_{N} )  =  \sum_{b }  \frac{N!}{N^{k/2}(N-b)!}  
	 \sum_{L(B)=b ; \atop B \in \mathcal{B}}  \prod_{x=1}^{N} \tr ( H_{B_{x}}  \tau_{1} ) .
\end{equation}
We proceed by showing that terms with $b < k/2$  and $b> k/2$ are either zero or decreasing with $N$, and so only the $b=k/2$ terms persist in the large $N$ limit.  

When $L(B) > k/2$ there must exist at least one bin $B_{x}$ that contains only 1 element, so $H_{B_{x}} = H_{j}$ for some $x$ and $j$.  This factor contributes $\tr ( H_{j} \tau_{1} )$ to the product, but by assumption $\tr ( H_{j} \tau_{1} )=0$ and so all such terms vanish.  As for the case with $L(B) < k/2$, we observe that as $N$ increases
\begin{equation}
	 \frac{N!}{N^{k/2}(N-L(B))!}  \rightarrow 0.
\end{equation}
Furthermore, for all $N>k$ the factor
\begin{equation}
 	 \sum_{L(B)= b ; \atop B\in \mathcal{B}}  \prod_{x=1}^{N} \tr( H_{B_{x}}  \tau_{1} ) ,
\end{equation}
is constant with $N$ as the number of canonical partitions stops increasing.  Therefore, for any $b< k/2$ the product of these terms vanishes with $N$.

This leaves only $b=k/2$ terms as potentially non-vanishing. Note that, if $k$ is an odd number there are no suitable integer $b$ values and so all odd moments will vanish with increasing $N$.  Assuming $k$ is even, the only non-vanishing partitions consists of pairings, such that each bin contains either 2 elements or none.   That is, non-vanishing $B$ have $H_{B_{x}}=H_{j}H_{k}$ or $H_{B_{x}}=\id$ for all $x$.  Putting these results together we have
\begin{eqnarray}
 	\lim_{N \rightarrow \infty}	\tr(  \hat{Q}^{k} \tau_{N} ) & = & \left( \lim_{N \rightarrow \infty} \frac{N!}{N^{k/2}(N-k/2)!} \right) \nonumber
 	\\ & & \times \sum_{B \in \mathcal{B}_{\mathrm{pair}}} \tr( H_{B_{x}}  \tau_{1} ) ,
\end{eqnarray}
where $ \mathcal{B}_{\mathrm{pair}}$ is the set of canonical pairings. The expectation value only depends on the $2^{\mathrm{nd}}$ moments of $\tau_{1}$ and so we can replace $\tau_{1}$ with the Gaussian state with the same second moments, namely $\tau_{\infty}$.  The combinatorial factor approaches $1$ and so
\begin{equation}
 \lim_{N \rightarrow \infty}	\tr ( H^{(k)} \tau_{N} )  = \sum_{B \in \mathcal{B}_{\mathrm{pair}}} \tr ( H_{B_{x}}  \tau_{\infty} )  = 	\tr ( H^{(k)} \tau_{\infty} ) .
 \end{equation}
In the simple case where the moment is a product of identical factors, so $H^{(k)}=H^{k}$, we have
\begin{eqnarray}
\label{EqWick}
 \lim_{N \rightarrow \infty}	\tr (  H^{k} \tau_{N} ) & = &| \mathcal{B}_{\mathrm{pair}}| \tr ( H^{2}  \tau_{\infty} )^{k/2} .
 \end{eqnarray}
 The number of canonical (unordered) pairings of $k$ numbers is simply $|\mathcal{B}_{\mathrm{pair}} | = (k-1)(k-3) \cdots 1$, which is known as a double factorial $(k-1)!!$. 
Consider the above results for when the input state is Gaussian, and so unchanging.  This tells us that the higher moments of a Gaussian state are determined its second moments, as captured by Eq.~(\ref{EqWick}), which is a well-known result called Wick's theorem.
 
\section{Matrix element convergence}
\label{APPMatrixElements}

This appendix provides a proof of Thm.~\ref{pointwiseMatrix}.  We move from statements about characteristic functions to operators by recalling that for an operator $B=\kb{\psi_{j}}{\psi_{k}}$ acting on an $m$-mode Hilbert space we have 
\begin{equation}	
\label{purity}
	 \tr ( B \tau  ) =(2 \pi)^{-m} \int  \chi_{B}(\vec{r}) \chi_{ \tau}(\vec{r}) d \vec{r} .
\end{equation}
Similar reasoning allows use to deduce that since $\tr ( BB^{\dagger} )=1$ and $\tr ( \tau \tau^{\dagger} )  \leq 1$, we know
\begin{eqnarray}
	(2 \pi)^{-m} \int  |\chi_{B}(\vec{r})|^{2}  &=& 1, \\ \nonumber
	 	(2 \pi)^{-m} \int  |\chi_{\varphi}(\vec{r})|^{2} & \leq& 1.
\end{eqnarray}
The absolute difference in expectation values between $\tau_{N}$ and $\tau_{\infty}$ is
\begin{eqnarray}
	D_{N} & = & |  \tr (  B \tau_{N}   - \tr ( B \tau_{\infty} ) | , \\ \nonumber
	& = &\frac{1}{(2 \pi)^{m}}  \int  \chi_{B}(\vec{r})\left[ \chi_{\tau_{N}}(\vec{r}) - \chi_{\tau_{\infty}}(\vec{r})   \right] d \vec{r} , \\ \nonumber
	& = &\frac{1}{(2 \pi)^{m}} \int  \chi_{B}(\vec{r}) \Delta_{N}(\vec{r})d \vec{r} ,
\end{eqnarray}
where $\Delta_{N}=\chi_{\tau_{N}}(\vec{r}) - \chi_{\tau_{\infty}}(\vec{r})   $. The proof proceeds by splitting the integral up into two parts so $D_{N}=D'_{N}+ D''_{N}$. We take $D'_{N}$ to be an integral over a large but finite ball of radius $R$ and $D''_{N}$  over the complement.   Over the complement we have that
\begin{eqnarray}
	D''_{N} & = &  \frac{1}{(2 \pi)^{m}}  \int_{|\vec{r}|>R}  \chi_{B}(\vec{r}) \Delta_{N}(\vec{r}) d \vec{r} , \\ \nonumber
	| D''_{N} | & \leq & \frac{1}{(2 \pi)^{m}} \left(  \int_{|\vec{r}|>R}  | \chi_{B}(\vec{r}) |^{2} d\vec{r} . \int_{|\vec{r}|>R} |\Delta_{N}(\vec{r})|^{2} d \vec{r} \right)^{\frac{1}{2}}, \\ \nonumber
\end{eqnarray}
where we have used the Cauchy-Schwarz inequality. From Eq.~(\ref{purity}) we can know $\int   | \chi_{B}(\vec{r}) |^{2} =1$ and so the integral over  $|\vec{r}|>R$ can be made arbitrarily small by increasing $R$.  Formally, for any $\epsilon'>0$ we can find an $R$ such that $\int_{|\vec{r}|>R}   | \chi_{B}(\vec{r}) |^{2}  \leq  \epsilon' $.  Furthermore,  Eq.~(\ref{purity}) entails that the integration over $|\Delta_{N}(\vec{r})|^{2}$ must be less than 2. Hence, we deduce
\begin{eqnarray}
	| D''_{N} | & \leq & \frac{\left( 2  \epsilon' \right)^{\frac{1}{2}}}{ (2\pi)^{m}} , \\ \nonumber
\end{eqnarray}
which holds for all $N$.  As for the integral inside radius $R$ we have 
\begin{eqnarray}
	| D'_{N} | & \leq & (2 \pi)^{-m} \int_{|\vec{r}| \leq R} | \chi_{B}(\vec{r})  \Delta_{N}(\vec{r}) | d \vec{r} . \\ \nonumber
\end{eqnarray}
For all characteristic functions $| \chi_{B}(\vec{r}) | \leq \tr ( \sqrt{B^{\dagger}B} )$ and so for $B=\kb{\psi_{j}}{\psi_{k}}$ we have $| \chi_{B}(\vec{r}) | \leq 1$.  Furthermore we know that within a finite ball $\Delta_{N}(\vec{r})$ vanishes uniformly, so for any $\epsilon'>0$ there is a $N_{\epsilon}$ such that for all $N>N_{\epsilon'}$ we have
\begin{eqnarray}
	| D'_{N} | & \leq & (2 \pi)^{-m} \int_{|\vec{r}|<R} \epsilon' d \vec{r} = \frac{\epsilon'' V}{(2\pi)^{m}} , \\ \nonumber
\end{eqnarray}
where $V$ is the volume of the ball.  Combining these results we have, for $N>N_{\epsilon'}$, that
\begin{equation}
	| \bra{\psi_{k}} \tau_{N} \ket{\psi_{j}} - 	\bra{\psi_{k}} \tau_{\infty} \ket{\psi_{j}} | < \frac{\left( 2  \epsilon' \right)^{\frac{1}{2}}+\epsilon'' V}{ (2\pi)^{m}}.
\end{equation}
Since $\epsilon'$ and $\epsilon''$ can be made arbitrarily small, we have proven Thm.\ \ref{pointwiseMatrix}.

\end{document}